\documentclass[12pt]{article}
\usepackage{euscript,amsmath,amssymb,epsfig,latexsym}

 \newcommand{\be}{\begin{equation}}
\newcommand{\ee}{\end{equation}}
\newcommand{\bea}{\begin{eqnarray}}
\newcommand{\eea}{\end{eqnarray}}

\def\CQG {Class. Quant. Grav.}
\begin{document}
\title{Catalysis of Black Holes/Wormholes Formation  in High Energy
Collisions\footnote{Extended version of the talk at the International Bogoliubov Conference
$"$Problems of Theoretical and Mathematical Physics$"$, Moscow-Dubna, August 21-27, 2009
}}
\author{I.Ya. Aref'eva,\\
\small{\it Steklov Mathematical Institute, Russian Academy of Sciences,}\\
\small{\it Gubkina str. 8, 119991, Moscow, Russia}}
\date {~}
\maketitle

\begin{abstract} We discuss various mechanisms of catalysis of black holes/wormholes (BH/WH) formation in
particles collisions. The current paradigm suggests that BH/WH formation in
particles collisions will happen
when   center of mass energies of colliding particles is sufficiently above
the Planck scale (the transplanckian region).

To estimate the BH/WH production we use the classical geometrical
cross section. We confirm the classical geometrical cross section of
the BH production reconsidering the process of two transplanckian
particles collision  in the rest frame of one of incident particles.
This consideration permits to use the standard Thorne's hoop
conjecture for a  matter compressed into a region to prove a variant
of the conjecture dealing with a total amount of
compressed energy in the case of colliding particles.

We calculate geometrical
cross sections for different processes and for different
background, in particular, for (A)dS.
We show that results are in agreement with closed trapped surface (CTS)  estimations
though   there are
 no general theorems
providing  that the BH formation follows from CTS's formation.

We show that the process of BH formation is catalyzed by the  negative cosmological
constant and by a particular scalar matter, namely  dilaton,
while it is relaxed by the positive cosmological constant and at a  critical
value just turns off.
Also we note that
the cross section is sensible  to the compactification
of extra dimensions and to the particular brane model.
\end{abstract}

\newpage

\section{Introduction}

Gravity  does not play
a role in the usual high energy terrestrial physics. However,  in the TeV gravity scenario \cite{AADD} the processes with
energy about few TeV become transplanckian  and  the
gravity is important.

Black holes formation in  collisions of transplanckian  particles is
one of outstanding problems in theoretical physics. Our aim in this
talk is to overview the current understanding of the problem.

Study of transplanckian  collisions in gravity has a long history.
In 80's-90's  the problem has been discussed mainly in  superstring theory
frameworks \cite{thooft_s,ACV87,GM87,Soldate,ACV89,ACV90,ACV93,AVV94,AVV95}
and was considered as an
academical one, since the four dimensional Planck scale  $E_{Pl}$ is
$\approx 10^{19}$ GeV, and   energies satisfying $\sqrt{s}>E_{Pl}$ wholly
out of reach of terrestrial experiments.

The situation has been changed after the proposal of TeV
gravity scenario \cite{AADD}. The
D-dimensional Planck energy $E_{Pl,D}$ plays the fundamental role in TeV gravity, it  has the electroweak
scale of $\sim$ TeV, as this would solve the hierarchy problem. TeV
gravity  is strong enough to play a role in elementary particle
collisions at accessible energies.

The TeV gravity assumes the brane world scenario \cite{VR-MS} that means  that all
light
particles
(except gravity) are confined to a brane with the 4-dimensional world
sheet  embedded  in the $D$-dimensional  bulk. The collider
signatures of such braneworld scenarios would be energy
non-conservation due to produced gravitons escaping into the bulk,
signatures of  new Kaluza-Klein particles   as well as signatures of
black hole (BH) \cite{BF,IA,DL,ADMR98,GidThom} and more complicated objects
such as wormholes
(WH) formations \cite{AV-TM,MMT,NS} (see \cite{INov,NovNov} about WHs in astrophysics).

According the common current opinion  the process of BH formation
in transplanckian collision of particle may be adequately described
using classical general relativity. We also believe that the same is true for the WH
production \cite {AV-TM}. Calculations based on classical general
relativity support \cite{penrose,eardley} the simple geometrical cross section of
black hole production in particles collisions, which is proportional to the area of the disk
\be
\label{g-cr-f}
\sigma = f \pi R_S^2(E),
\ee
 where $R_S$ is
the Schwarzschild radius of the black hole formed in the particles
scattering process and it is defined by the center-of-mass collision energy
$E=\sqrt{ s}$, and $f$ is a formation factor of order unity.
Colliding particles in
hadron colliders are  partons  and the total cross section for
black hole production  is calculated using a factorization hypothesis in
which the parton-level process is integrated over the parton density
functions of the protons \cite{PDF}.
If  the  geometric cross section were true and colliding particles carry few TeV,
the LHC
would produce black holes at a rate $\sim$1 Hz for $M_{Pl,D}=1$
TeV, becoming a black hole factory \cite{DL,GidThom}.

However, BH formation in particle collision is a threshold phenomena
 and the threshold is of
order the
Planck scale $M_{Pl,D}$\cite{gringrich07}. The exact value of the threshold is
unknown since it depends of  quantum gravity
 description of
colliding particles. BH production rates depend on the value of
$M_{Pl,D}$ \cite{PDG}.
Current bounds\cite{Duperrin}
are dimension-dependent but lie around
$
M_{Pl,D} \gtrsim 1\, {\rm TeV}$.
Taking   simple estimation for cross section (\ref{g-cr-f}) with $f\sim 1$
above the threshold one can conclude that
 the cross section of semiclassical
BHs production  above the threshold at the LHC
varies between 15 nb and 1 nb for the Planck scale between 1 TeV and 5 Tev.
Note, that this cross section is compatible, for example, with $t\bar t$
production \cite{Lansberg08}.
Just after production BHs quickly ($\sim 10^{-26}$~s)
evaporate via Hawking radiation~\cite{Hawking75} with a characteristic
temperature of $\sim 100$~GeV~\cite{DL,GidThom}. However, since  produced BHs
are light they decay
into only a few high energy particles and this would be  difficult to
disentangle from the background \cite{DvaliSibir}.

A natural question arises: can we   catalyze the semiclassical process
of the BHs formation and increase the production factor $f$ in (\ref{g-cr-f})?
This is the main question that we are
going to discuss  in this talk. Let us note that in this talk we are going
to deal with
semiclassical consideration and  make few notes of the region of its
applicability. We will search for  theoretical  possibilities to
increase the formation factor in the formula for the geometrical cross section.
There are effects that  work  in the opposite
direction and  push the
collision energy needed for BH formation considerably higher than
$M_{Pl,D}$. These effects are related with
the energy loss by colliding particles prior to the formation of the
BH horizon \cite{Meade:2007sz} and the effects of the charge
\cite{Yoshino:2006dp}.

In fact there are  few possibilities
in our disposal to increase the formation factor. We are going to explore the following
proposals
\begin{itemize}
\item  find  effects related with nontrivial dynamics of 3-brane embedding
in D-dimensional space-time
\item
change
 the background (4-dimensional background or D-dimensional one), in particular, we
 can to add 4-dimensional
the cosmological constant (or cosmological constant in D-dimensional
space-time)

\item take into account that shock wave in D-dimensional space-time
can be made of from closed string excitations.
\end{itemize}

Some attempts toward these directions are presented in this talk.

In the last years numerous papers have been devoted to
improvement  calculations based on classical general relativity
to get more precise estimates for the cross-section (\ref{g-cr-f})
\cite{Yoshino:2006dp,Yoshino05a}. In particular, numerical calculations have been performed and they  confirmed
 (\ref{g-cr-f}) and gave
the estimations for the production factor \cite{Choptiuk}.
The effects of finite size\cite{Kohlprath,Giddings04}, charge\cite{Yoshino:2006dp}
and spin\cite{Yoshino07} have been
considered.
It has been found that the effects of mass, spin, charge and finite size of the
incoming particles are rather small. The effects of the cosmological constant have
been considered in papers\cite{Gubser08,Alvarez08,0803.3226,Gubser09,Shuryak09,0902.3046,ABG,
IA-AB-TMP,ABJ}. In these papers estimation of the cross section of the BH production
 for particles colliding in $(A)dS$
backgrounds have been made. AdS case has been studied mainly
within the AdS/CFT context, and dS case within possible cosmological applications
\cite{ABG,ABJ}. It has been found that the  negative cosmological constant
increases  the cross section, meanwhile the positive cosmological constant works
in the opposite direction destroying the trapped surface at the critical value of
the cosmological constant and by this reason presumably  holding up
the BHs production.

Quantum field theory is a local theory in the Minkowsky space \cite{BS,BLOT}.
However if we take into account effects of quantum gravity then some form of nonlocality is occurred.
The problem of (non)locality in quantum gravity was addressed in \cite{thooft_s,p-adic,GM87},
and more recently in \cite{Giddings:2009ae}

The talk is organized as follows. In Sect. 2 we present a setup to
study the BH formation in particle collisions.
 We follow \cite{AVV94} and present a natural
generalization of this approach to the brane world case.
We briefly summarize the main achievements of study the BH formation in flat
background within the classical general gravity picture
(trapped surfaces technics) and discuss why
we can trust classical description.

In Sect.4. we discuss a  physical picture of a black hole
formation in the rest frame of one of colliding particles.
This picture permits from simple calculations make a conclusion
about the BH formation.

In Sect.  5  we generalize the above consideration to the case
of the BH formation in (A)dS. This consideration presents a special interest
because in contrast to the flat background in the AdS case there are no general
theorems that guarantee the BH formation from the CTS formation.
 We discuss here an influence of
structure of the shock wave on the  BH formation. This structure
is defined by gravitation interaction of
matter fields.
We also
perform
 calculations of the cross section of the BH production in different cases
 and  show  agreements with    trapped
surfaces results. We conclude that the process of BH formation
 is catalyzed by a negative cosmological
constant as well as  by the string dilaton.

\newpage
\section{Setup. Milestones and
Notations}
\subsection{D-dimensional Planckian Energy }

\begin{figure}[h!]
\centering
\includegraphics[width=5cm]{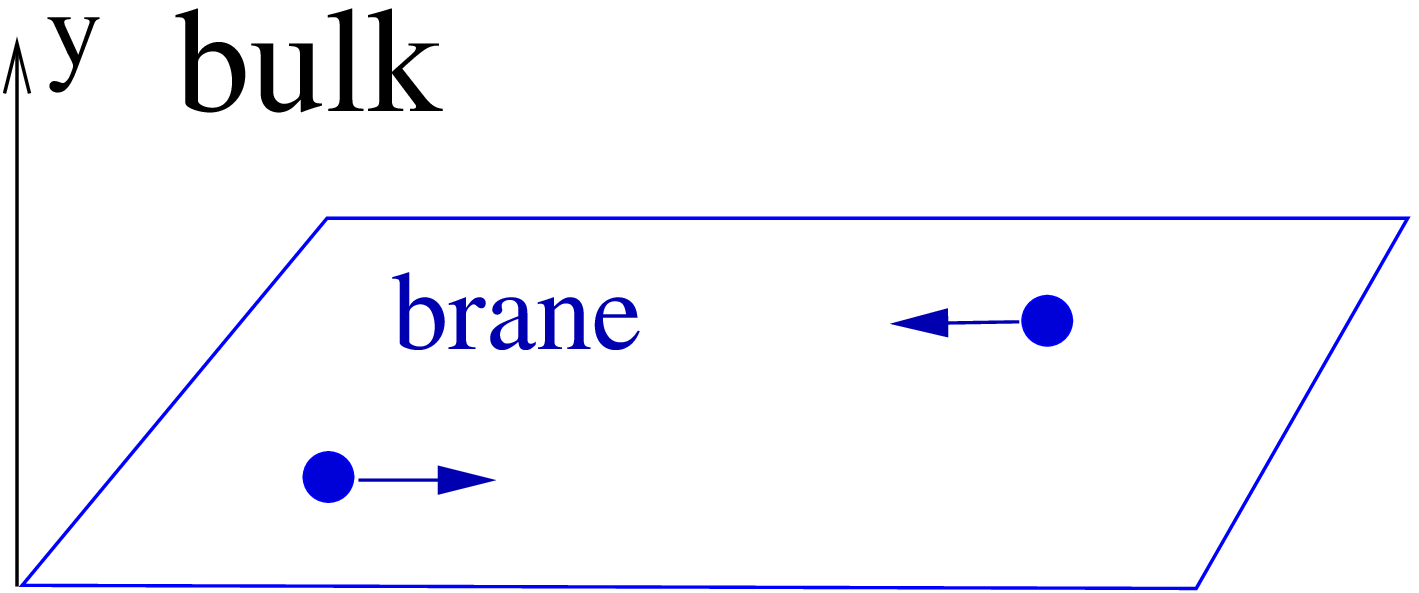}$A\,\,\,\,\,\,\,\,\,\,$
$\,\,\,\,\,\,\,\,\,\,$
\includegraphics[width=5cm]{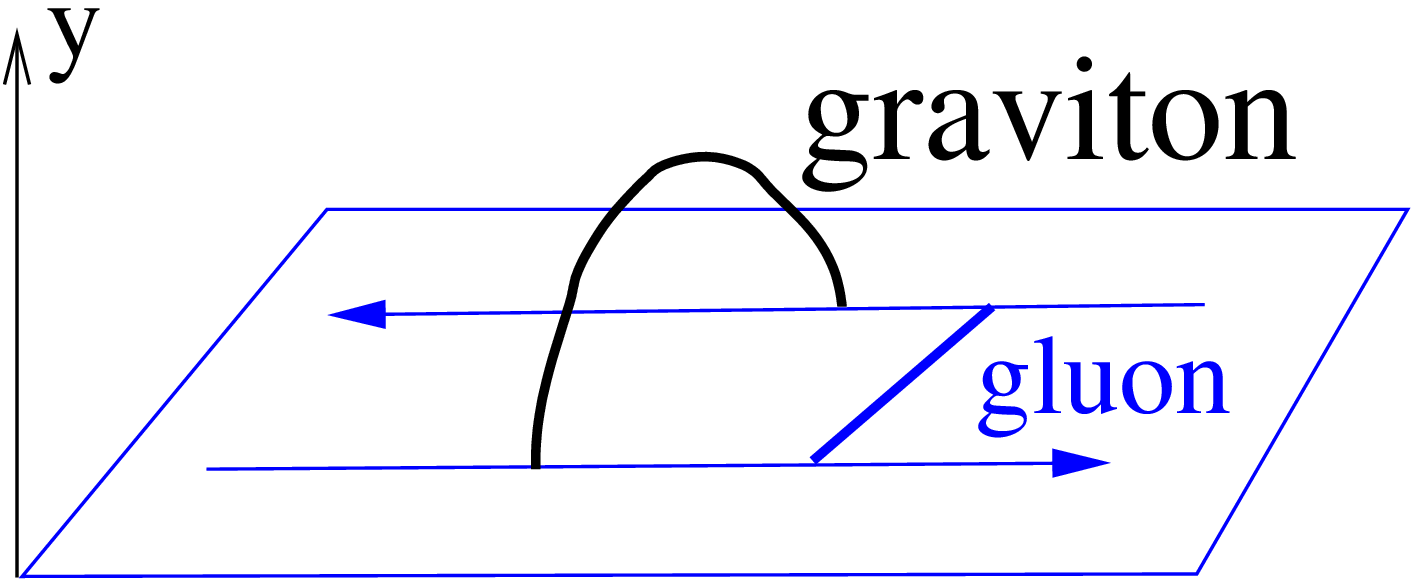}
$B$
\caption{A. Colliding particles on the brane. B. D-dimensional graviton
and 4-dimensional gluon  exchanges.}\label{Fig:1}
\end{figure}

In TeV gravity scenario we assume that all particles and fields (except gravity)
are confined to a brane with 4-dimensional world sheet  embedded  in the
$D$-dimensional  bulk. Matter fields leave on the brane and do not feel
extra dimensions, only gravity feels $n=D-4$
extra dimensions, see Fig.\ref{Fig:1}.

$$\,$$
$$\,$$

According the common current opinion  the process of BH
formation in transplanckian collision of particle,
i.e. in regions where
\be \sqrt{s}>>E_{Pl},
\label{TP}\ee
 may be
adequately described using classical general relativity. We also believe that the same concerns
the WH production.
This is because
in the transplanckian region (\ref{TP})  the de Broglie wavelength  of a particle
\be
l_{{\cal B}}=\frac{\hbar c}{E}
\ee
is less than the Schwarzschild radius corresponding to this particle,
\be
\label{lc-RS}
l_{{\cal B}}<<R_{S,D},
\ee
here $R_{S,D}$ is the D-dimensional Schwarzschild radius in TeV gravity.
In phenomenologically reasonable models with $n\geq 2$ the Schwarzschild radius
corresponding to  colliding particles
with energies $\approx$ 1 TeV  is  $R_{S,D}\gtrsim 10^{-16}cm$.
In the usual 4-dimensional gravity the  Schwarzschild radius corresponding
to the same particles is
of order $R_{S,4}\sim 10^{-49}cm$, that is a negligible quantity
comparing  with the de Broglie wavelength  of  particles with energy about few TeV.

Although these type of processes are classical it is instructive to have a full
picture starting from a general quantum field theory  setup and pass explicitly
to the classical description of processes in question.
This point of view is useful  to deal with effects on the boundary of the classical
 applicability.
By this raison we start in the next subsection from this general setup
\cite{AVV94},
and in Sect.2.? we present the brane extension of this
approach \cite{IA}.

\subsection{Transition amplitudes and cross section of the BH/WH production}

We start from  quantum mechanical formula for the
cross section $
\sigma_{AB}$ of
a process
\be
|\,A>\,\,\,\Rightarrow\,\,|\,B>.
\ee
 To calculate this cross section we  calculate  the transition amplitude
between these states
\be
\label{FI}
<A|\,B>=\int \Psi ^*_A(X_A,t){\cal K}(X_A, t;X_B,t^\prime)
\Psi _B(X_B,t^\prime)dX_AdX_B\ee
where $X$ are  generalized coordinates, specifying the system,
$\Psi _A(X,t)$,  is a wave function of the state $A$ including its asymptotical
dynamics.
The transition amplitude in the generalized coordinate representation
is given by the Feynman integral.
In our case we deal not only with particles but also with gravity.
In particular, we  discuss the process where the final state $|B>$
is the state corresponding to the black hole.
To this purpose we use a modification \cite{AVV94} of the standard formula
\cite{AFS}:

\begin{itemize}
\item For simplicity we  work in $1+3$ formalism where spacetime is presented
as  a set of slices
(more general formulation is described  in \cite{AVV94}). At the initial time $t$
we deal with a slice $\Sigma$ and at the final time $t^\prime$ with a slice $\Sigma^\prime$.

\item Generalized coordinates include a metric $g$ and matter fields $\phi$.
\item The state
at on a initial time is specified
by a  three-metric $h_{ij}$ and field $\phi $ and final state by a
three-metric $h_{ij}^\prime$ and  $\phi ^\prime$.
\item
 The transition amplitude in this generalized coordinate representation
is given by Feynman integral \cite{AVV94}

\be
 \label{FI-GR}
{\cal K}(h,\phi , t;h^\prime, \phi ^\prime,t^\prime)=\int e^{\frac{i}{\hbar}
S[g,\phi]}\prod
_{{\small{\small\begin{array}{ccc}&\phi |_{\tau =t}=\phi,\,\,
g |_{\tau =t}=h &\\&
\phi |_{\tau =t^\prime}=\phi ^\prime,\,\,
g |_{\tau =t^\prime}=h ^\prime &\\
\end{array}}}}~{\cal D}\phi (\tau){\cal D}g(\tau)
\ee
where the integral is over all four-geometries and field configurations
which match  given values on two spacelike surfaces, $\Sigma$ and $\Sigma^\prime$
and matter on them, $S[g,\phi]$ is the action. The integral in (\ref{FI-GR})
includes also summation over different topologies.

\item The transition amplitude given by the functional integral
includes gauge fixing and Faddeev-Popov ghosts (all these are omitted
in (\ref{FI-GR}) for simplicity).
\item We are interested in the process of a black hole creation in particles
collisions. Therefore,
\begin{itemize}
\item we
specify the initial configuration $h $ and $\phi  $ on $\Sigma $ without black holes,
i.e. causal geodesics starting from $\Sigma$ rich the future
null infinity
 \footnote{More precise,
 this condition  means that
$\Sigma $  is a partial Cauchy surface with
 asymptotically simple
past in a strongly asymptotically predictable space-time.};
\item
 we specify the final configuration $h^\prime$ and $\phi ^\prime$
on $\Sigma ^\prime$
as describing black hole, i.e. $\Sigma ^\prime$ contains a region from which the light
 does not rich the future null infinity
 \footnote{This  means that $\Sigma ^\prime$ is a partial Cauchy surface
 containing black hole(s), i.e.
 $\Sigma ^\prime-J^{-}({\cal T}^{+})$ is non empty.}.

\end{itemize}
The  explanation of  notions used in above footnotes is given in \cite{AVV94},
see also Appendix.
 For more details see \cite {HawkingEllis,Ward}.
\end{itemize}

\begin{figure}[h!]
\centering
\setlength{\unitlength}{1mm}
\begin{picture}(30,50)
\put(-20,0){
\includegraphics[width=4cm]{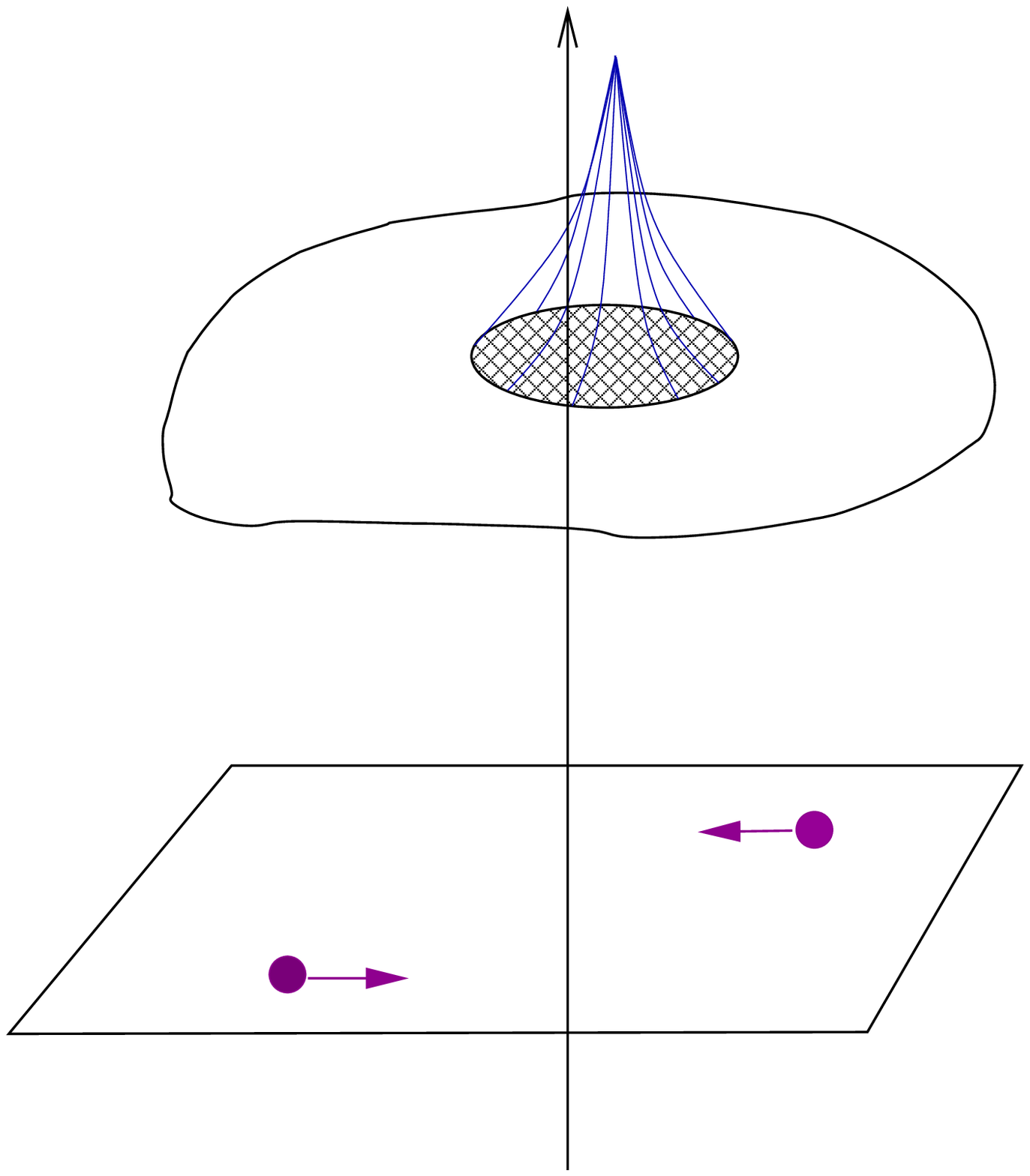}
}
\put(-10,26){\makebox(0,0)[lb]{$\Sigma^\prime$}}
\put(12,30){\makebox(0,0)[lb]{${\cal B}$}}
 \put(-13,8){\makebox(0,0)[lb]{$\Sigma$}}
\put(-5,42){\makebox(0,0)[lb]{$\tau$}}
\end{picture}
\caption{A slice  $\Sigma$ at $\tau =t$
is an initial slice with particles and  a slice $\Sigma ^\prime$ at $\tau =t^\prime$
 is a slice with a black hole ${\cal B}$. Null geodesics started from the shaded region
 do not reach null infinity.
 }\label{Fig:slice}
\end{figure}
In Figure \ref{Fig:slice}   a slice with two colliding particles at
 $\tau =t$, and $\tau =t^\prime$
with the  BH area are presented. To describe  such a process  in the framework of a general
approach  (\ref{FI-GR})  we have to find a classical solution of the Einstein
equations with the matter, our moving particles,  that corresponds to this picture,
Figure \ref{Fig:slice},
and then study quantum fluctuations. We do not have analytical solutions
describing this process.

Finding of such solutions is a very difficult
problem. It is solved only at low dimensional case, see
\cite{Hooft-3-dim-gr,3-dim-gr,IA} and
refs. therein.
 In 4-dimensional
case this problem has been solved numerically only recently by Choptiuk and Pretorius
\cite{Choptiuk}. The
solution, as it has been mentioned in Introduction, assumes a construction of a model for
gravitational particles. We present this construction    in the next subsection.

\subsection{D-dimensional gravitational model of relativistic particles}

To start a classical description of BH production in collision of elementary
particles we need a gravitational model of relativistic particles.
At large distances  the gravitational field of particle is the usual Newtonian field.
The simplest way to realiz this is just to take the exterior of the Schwarzschild
metric, i.e.
in D-dimensional case away a particle we expect to have
\be
\label{D-Sch}
ds^2= \left(1-(\frac{R_{S,D}}{R})^{D-3}\right)dt^2+
\left(1-(\frac{R_{S,D}}{R})^{D-3}\right)^{-1}dR^2+R^2d\Omega_{D-2} ^2,
\ee
where $R_{S,D}$ is the  Schwarzschild radius
\be
\label{sh-rad}
R^{D-3}_{S,D}(m)=\frac{16\pi G_Dm}{(D-2)\Omega_{D-2}}=\frac{2m}{(D-2)\Omega_{D-2}M^{D-2}_{Pl,d}},
\ee
 here $G_D$ is D-dimensional Newton
gravitational constant,
$c$ the speed
of light (in almost all formula we take $c=1$) and $\Omega _{D-2}$ is the
geometrical factor,
\be
\Omega_{D-2} = \frac{2\pi^{(D-1)/2}}{\Gamma[(D-1)/2]},
\ee
$\Gamma$ is Euler's Gamma function. Here we present D-dimensional formula,
 in particular for $D=4$,
$R_{S,4}(m)=2G_4m$. We also use the expression of the Schwarzschild radius
in therm of the Planck mass, $R_{S,4}(m)=m/4\pi M_4^2=m/ \bar{M}_4^2$.

The interior of the Schwarzschild metric is supposed to fill with some matter.
The simplest possibility is just to take a Tolman-Florides interior
incompressible perfect fluid solution
\cite{Tolman,Florides}.
 As another  model of  relativistic particles
one can consider a static spherical symmetric solitonic solution of
gravity-matter equations of motion,
the so-called  boson stars (authors of refs.\cite{bsrefs,Schunck:2003kk}
deal with 4-dimensional space-time, but it not a big deal to get D-dimensional extantions).

In the case of  brane scenario few comments  are in order.
In the simplest brane models we  deal with matter only on the brane and we do not have
matter out of the brane to fill the interior of the D-dimensional
Schwarzschild solution.
However, in the string scenario\footnote{Open string excitations are located on brane, closed string excitations
propagate on the bulk} there are closed string excitations which are suppose to
be available in the bulk.
One can assume that the matter in the bulk is a dilaton scalar field and deal
with string inspired D-dimensinal generalization boson stars
\be
\label{star}
ds^2= \left(1-(\frac{{\cal A}(R)}{R})^{D-3}\right)dt^2+
\left(1-(\frac{{\cal B}(R)}{R})^{D-3}\right)^{-1}dR^2+R^2d\Omega ^2,
\ee
where
\be
{\cal A}(R)=A+A_1/R+...\,\,.
\ee

Note, that stars in modified gravity and on  branes have been considered in
\cite{Babichev:2009fi} and \cite{Creek:2006je,0903.0066}, respectively.

\subsection{Shock wave as a model of ultra relativistic moving  particle}

\begin{figure}[h!]
\centering
\includegraphics[width=7cm]{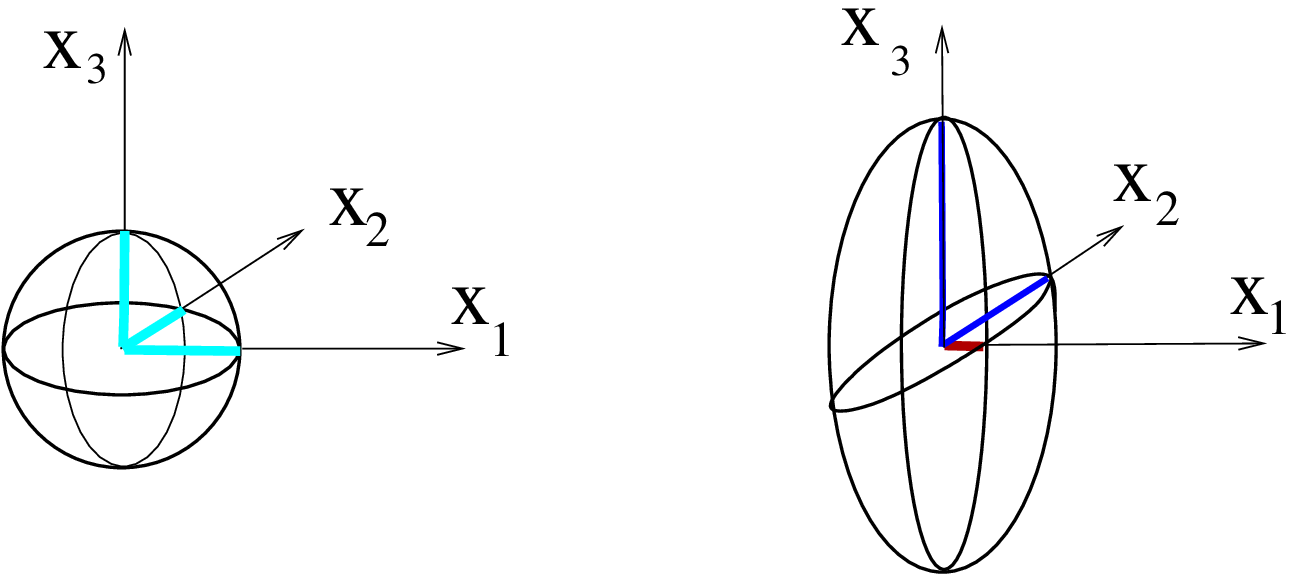}$A\,\,\,\,\,\,$
\includegraphics[width=3.5cm]{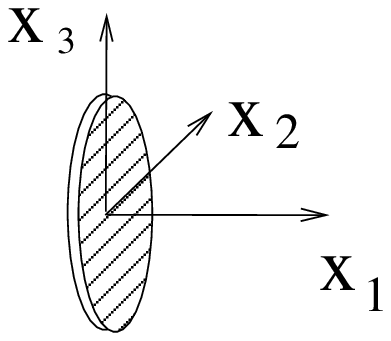}$\,\,\,\,B$
\caption{A.Flattening  of the  Schwarzschild sphere in the boosted coordinates.
 B. Schematic picture for the shock wave  as a flat disk.}\label{Fig:3}
\end{figure}

To consider ultra relativistic moving  particle we have to make a boost of
metric (\ref{star})
with the  large
 Lorentz boost factor $\gamma=1/\sqrt{1-v^2/c^2}$.
 The Schwarzschild sphere under this boost flattens  up to an ellipsoid, see Figure
 \ref{Fig:3}.

 $\,$

$\,$

  One can consider an approximation when
 $\gamma$ is taken infinitely large and
 $E=\gamma A$ is fixed.
 The result metric is the Aichelburg-Sexl
 (AS) metric \cite{AS71,thooft}, a gravitational
shock wave, where the non-trivial
geometry is confined to a $D-2$-dimensional plane traveling at the speed
of light, with Minkowski spacetime on either side,
\be
\label{AS}
 ds^2=-2dUdV+ dX_i^2+F(X)\delta(U)dU^2,\,\,\,\,i=2,3,..D-1,
 \ee
 where $V=(X^0+X^1)/\sqrt{2},\,\,U=(X^0-X^1)/\sqrt{2}$.
 The form of the profile of the shock wave $F$ depends on the behavior of
 ${\cal A}(r)$.

 In particular, in the infinite boost limit
where we also take $m \rightarrow 0$ and hold $p$ fixed, the
metric (\ref{D-Sch}) reduces to an exact shock wave metric (\ref{AS})
with the shape function $F$ being  the Green function of the $D-2$-dimensional Laplace equation
\be
\Delta_{R^{D-2}}F=-\frac{2p\sqrt{2}}{M_{Pl}^{D-2}}\delta^{(D-2)}(X).
\ee
here $\delta^{(D-2)}(X)=\prod^{D-1}_{i=2}\delta(X^i)$,
\be
\label{D-profile}
F(X)=\frac{p2\sqrt{2}}{(D-4)\Omega M_{Pl,D}^{D-2}}\frac{1}{\rho^{D-4}}\ee
where $\rho^2=(X^2)^2+...(X^{D-1})^2$. For D=4 the shape is
\be
F(X)=-\frac{p\sqrt{2}}{\pi M_{4}^2}\ln \frac{\rho}{\varepsilon}
\ee

\begin{figure}[h!]
\centering
\includegraphics[width=5cm]{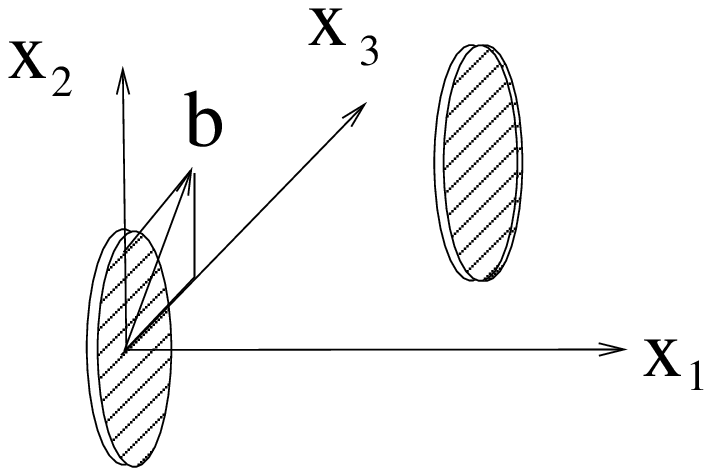}$A\,\,\,\,\,\,\,\,\,\,$
$\,\,\,\,\,\,\,\,\,\,$
\includegraphics[width=3cm]{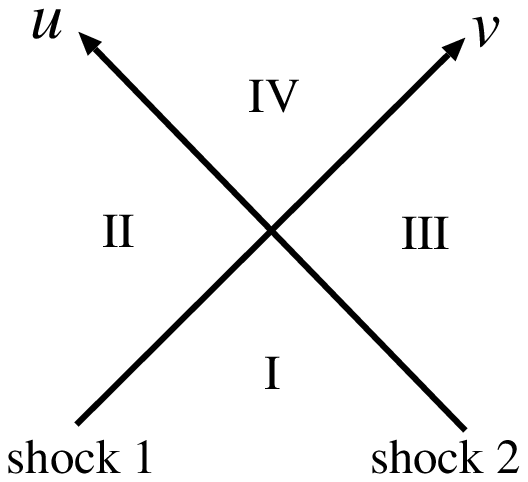}
$B$
\caption{A.  Ultra relativistic colliding particles in $D-1$-dimensional space;
 $b$ is the $D-2$-dimensional impact vector.
B. Ultra relativistic colliding particles in $U,V$-plane}\label{Fig:4m}
\end{figure}

Note, that  the metric (\ref{AS}) is obtained in the infinite boost limit when the source has
 zero rest mass.
For fast particles of
nonzero rest mass, the shock wave approximation breaks down far
away from the moving particle, more precisely  at transverse distances from
the source which are of the order of
\be
\label{sw-validity}
\ell \sim
{r_h(m)}/{\sqrt{1-{\rm v}^2}}.
\ee
 At these distances the field lines will spread
out of  the null transverse
surface orthogonal to the direction of motion.
But for ${ b}<<\ell$ one can use the shock wave field to extract the
information about the black hole formation to the leading order in
$m/p$.
 These shock wave are presented
in Figure \ref{Fig:4m} as sphere flattened up to the disk.
Two such shock waves, moving in opposite directions, see Figure \ref{Fig:4m}.B
 give the pre-collision geometry of the spacetime. Though the geometry
is not known to the future of the collision, since the shock wave
solutions inevitably break down when the fields of different
particles cross,  at the moment
of collision a trapped surface can be found~\cite{penrose,deathpayne,eardley}.

$\,$

$\,$

 According to \cite{penrose,deathpayne,eardley},
the trapped surfaces do form when ${ b} \lesssim R_{S,D}$, and have the
 area of the order $\sim R_{S,D}^2$, where $R_{S,D}$ is the horizon
radius given by (\ref{Thorn-conj-E}) (see below).

\begin{figure}[h!]
\centering
\setlength{\unitlength}{1mm}
\includegraphics[width=4cm]{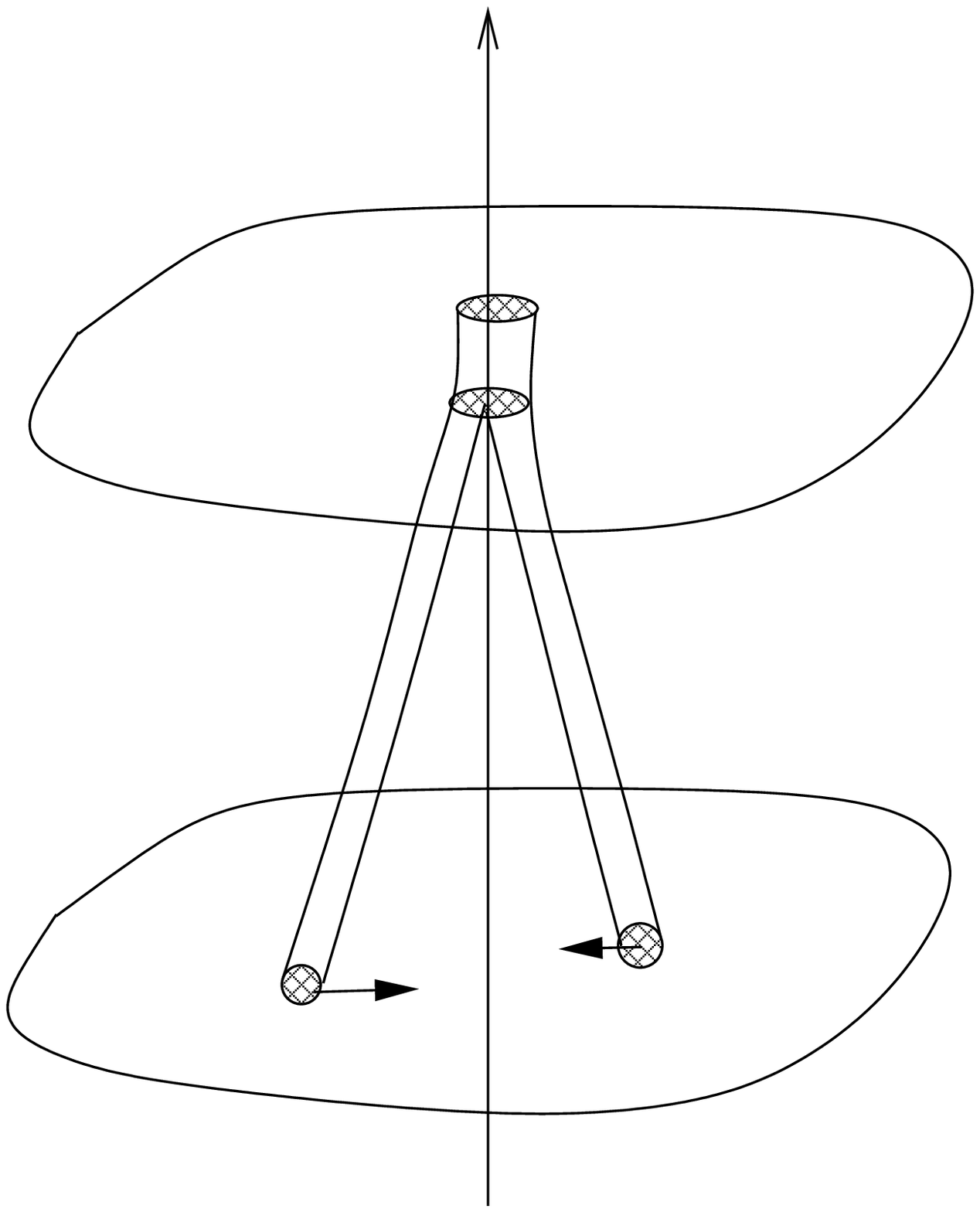}
$A\,\,\,\,\,\,\,\,\,\,$
\includegraphics[width=4cm]{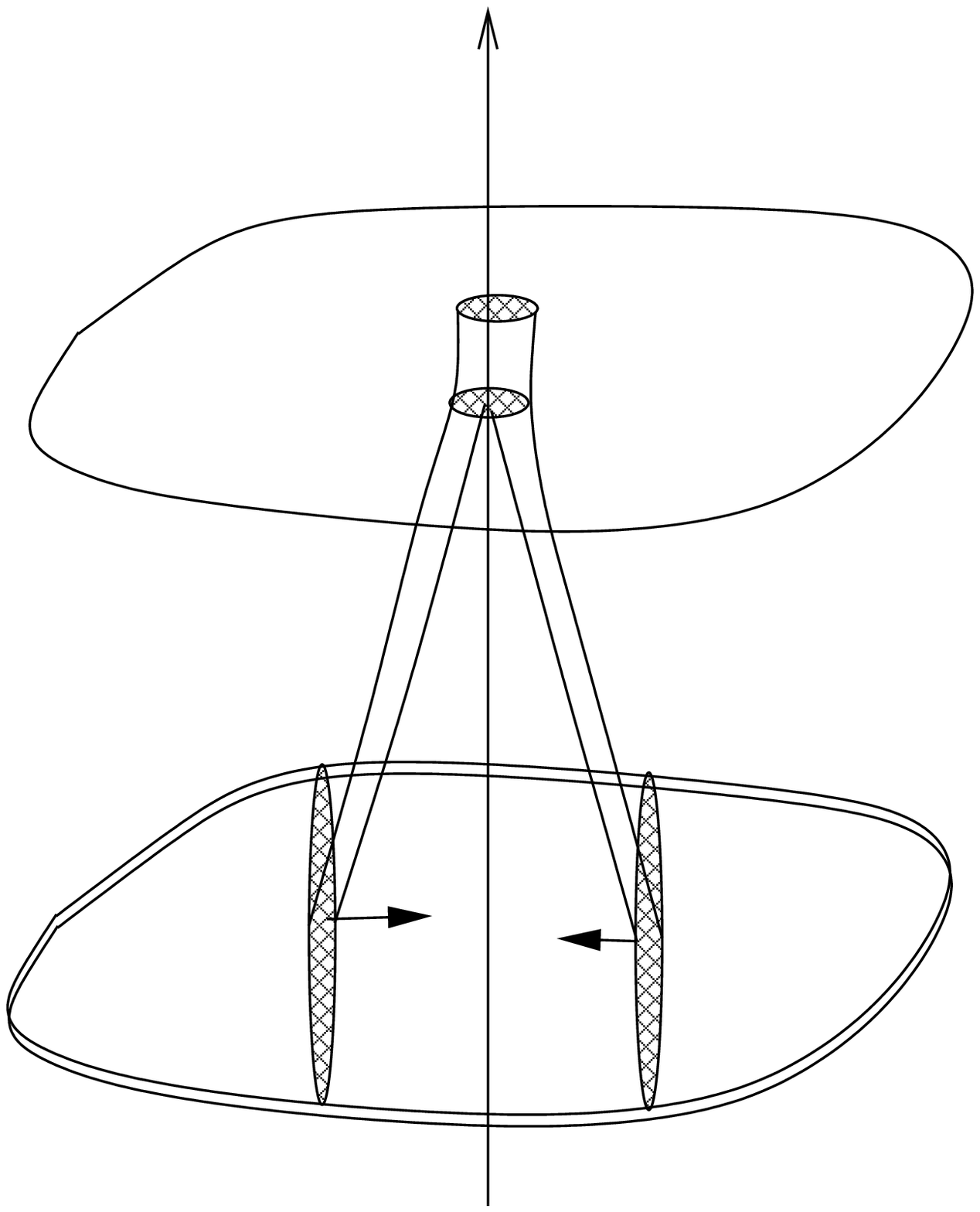}$B$
\caption{A. Colliding two stars; the initial space-time is asymptotically flat.  B
  Colliding shock waves as models of ultra relativistic  particles;
  the initial space time is not
  asymptotically flat.
 }\label{Fig:col}
\end{figure}

Infinitely thin
shock is an  idealization. In reality the shocks will have a finite
width $w$ since $\gamma$
is large but not infinite. The corresponding  shocks have width
$
w_{class}\sim r/\gamma,
$
depending on the transverse distance $r$. Infinitely thin idealization leads  to
an appearance in the intersection of the planes
of the two shock-waves  a divergent curvature invariant \cite{SR0401116}.
In \cite{RG0409131} has  shown that this problem is an artifact of the
unphysical classical point-particle limit and  for a particle
described by a quantum wavepacket, or for a continuous matter
distribution, trapped surfaces indeed form in a controlled regime.

\subsection{D-dimensional Thorne hoop conjecture and geometrical cross section}
\label{2.5}
We expect to get the BH formation due to nonlinear  interaction of gravitational
fields produced by particles.
The BH formation in classical general relativity is controlled by the Thorne hoop
conjecture \cite{thorne}. According the D-dimensional  version of this conjecture
 if a total amount of matter mass $M$ is compressed
into a spherical region of radius $R$, a black hole will form if
$R$ is less than the corresponding Schwarzschild radius
\be
\label{Thorn-conj}
R<R_{S,D}(M),
\ee
here $R_{S,D}(M)$  is given by (\ref{sh-rad}).

In the case of ultra relativistic particle collisions   the main argument
for black hole formation is
 based on a modification of Thorne's hoop
conjecture. According this modified conjecture  if a total amount of energy $E$ is compressed
into a spherical region of radius $R$, a black hole will form if
$R$ is less than the corresponding Schwarzschild radius
\be
\label{Thorn-conj-E}
R<R_{S,D}(E)\equiv \left(  \Omega _n\,\frac{G_D E}{c^4}
\right)^{\frac{1}{n+1}},
\ee
Note that in this modified conjecture the horizon radius $R_{S,D}$
is set by the center-of-mass collision energy
$E=\sqrt{ s}$.

Few remarks are in order concerning this formulation. Literally speaking,
as it is formulated above,
it is not applicable in all situations. But this conjecture does applicable
for two colliding particles. There are several calculations and arguments
supporting this conjecture:
\begin{itemize}
\item
One set of arguments is related with
examining trapped surfaces formation in collisions
of ultra relativistic particles \cite{penrose,GidThom}.
Note that commonly used  evidence for black hole formation in collision of particles
comes from the
study of the collision of two  Aichelburg-Sexl shock wave.
This argument  assumes that there is a solution
interpolating between two shock waves and BH, Figure \ref{Fig:col}.B. However
with this argument  there is a problem that
a space time with a shock wave  is not asymptotically flat, that assumed in our scheme
\footnote{The AS metric also has a naked singularity at the origin.
This is considered as an artifact of having used a black hole
metric as the starting point, and assumed to be removed by taking a suitable mass
distribution.}
\item The same problem is also with colliding plane wave \cite{AVV94}.
An advantage to deal with plane waves is that
in this case one can construct explicitly the metric in the interacting region.

\item There is a non-trivial possibility to reduce the proof
Thorne's hoop
conjecture for ultra colliding relativistic particles to
Thorne's hoop
conjecture for  slow  moving relativistic particles (see Sect. 3 below)
\item There are resent numerical calculations supporting (\ref{Thorn-conj-E})
\cite{Choptiuk}. In \cite{Choptiuk} as a model of particles
the boson star is taken \cite{bsrefs}. Choptiuk and Pretorius have got a remarkable
result that black holes  do form at high velocities in
boson star collisions and they found also that this happens
 already at a $\gamma$-factor of roughly  one-third
predicted by the hoop conjecture.
\end{itemize}

On the modified Thorne's hoop
conjecture for ultra colliding relativistic particles
 the so-called geometrical cross section of BH production is based.
It estimates the black hole
production cross section by the horizon area of a black hole whose
horizon radius $R_{S,D}$ is set by the center-of-mass collision energy
$E=\sqrt{ s}$, eq. (\ref{Thorn-conj-E}).
\begin{figure}[h!]
\centering
\includegraphics[width=5cm]{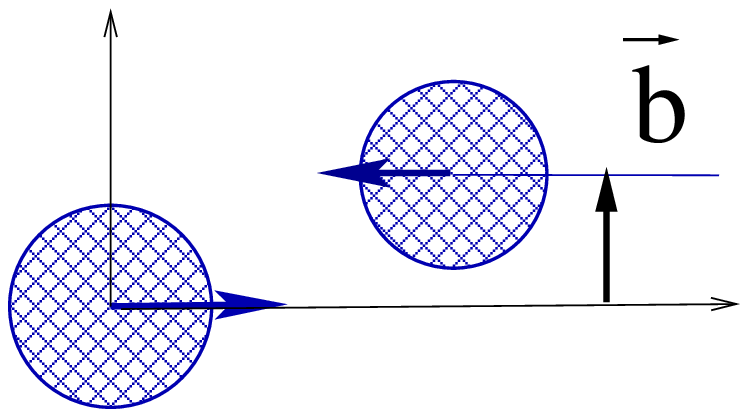}$A\,\,\,\,\,\,\,\,\,\,$
$\,\,\,\,\,\,\,\,\,\,$
\includegraphics[width=6cm]{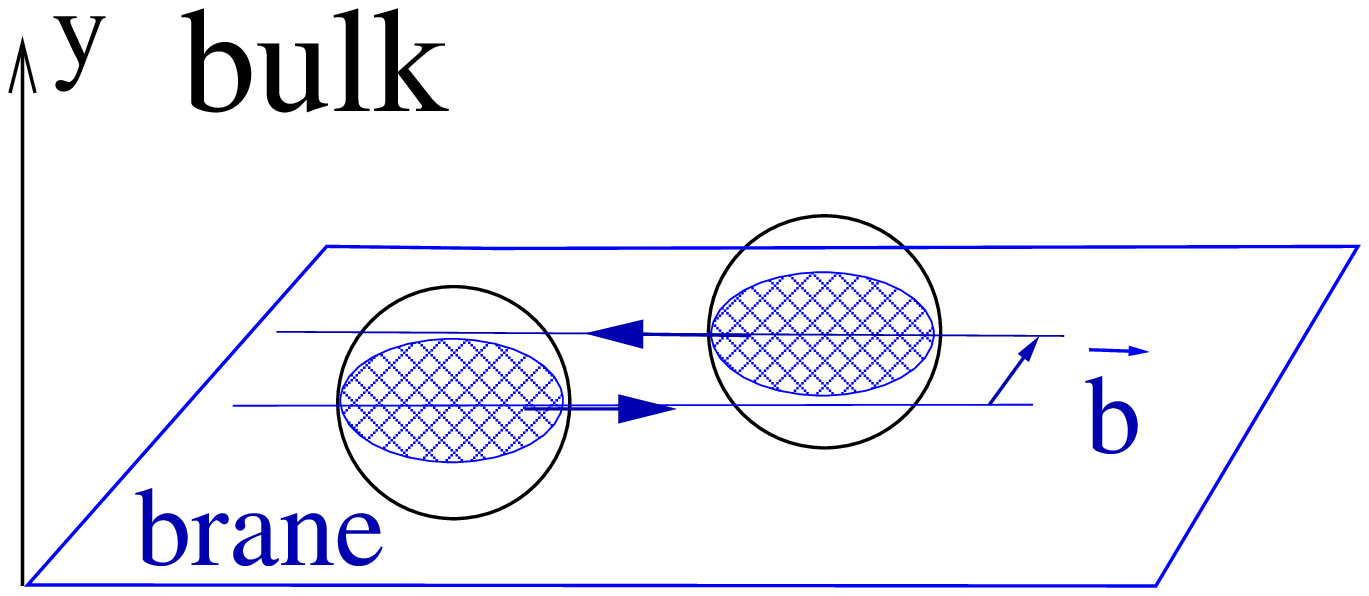}
$B$
\caption{A.  Colliding particles in $D-1$-dimensional space:
 $b$ is $D-2$-dimensional impact vector
and $\sigma \approx {\cal D} R_{S,D}^{D-2}$.
B. Colliding particles on the $3$-brane: 2-dimensional impact vector $b$
and
$\sigma \approx \pi  R_{S,D}^{2}$. A shaded region indicates the projection
of $D-1$ dimensional Schwarzschild sphere onto  the 3-brane}
\label{Fig:6}
\end{figure}
 This estimation assumes  that when the impact parameter
${ b}$ is smaller than $R_{S,D}$
then the probability of formation of a black hole is
close to $1$,
\be
\label{geom-cross}
\sigma_{BH,D}\approx {\cal D}_{D-2} R_{S,D}^{D-2}(E),
 \ee
${\cal D}_{D-2}$ is the volume of a plane cross section of the $D-2$
 dimensional unit sphere, see  Figure \ref{Fig:6}.A where $b$ is $D-2$-dimensional vector, i.e
 the area of of $D-2$-dimensional disk,
 \be
 \label{D-Disk}
 {\cal D}_{D}=\frac{\pi ^{D/2}}{\Gamma(1+\frac{D}{2})};
 \ee
In the  4-dimensional case this estimation gives

 \be
\sigma_{BH,4}\approx \pi R_{S,4}^2(E)
  \ee

For the 3-brane embedding in the D-dimensional space-time we have
 \be
\sigma_{BH,3-brane}\approx \pi R_{S,D}^2(E)
  \ee
  since our particles are restricted on the
 3-dimensional brane and the impact vector $b$ is two dimensional vector,  see Figure
\ref{Fig:6}.B.

\subsection{Looking from the rest frame of one of the
incident particles}
\begin{figure}[h!]
\centering
\setlength{\unitlength}{1mm}
\begin{picture}(100,50)
\put(-15,0){
\includegraphics[width=8cm]{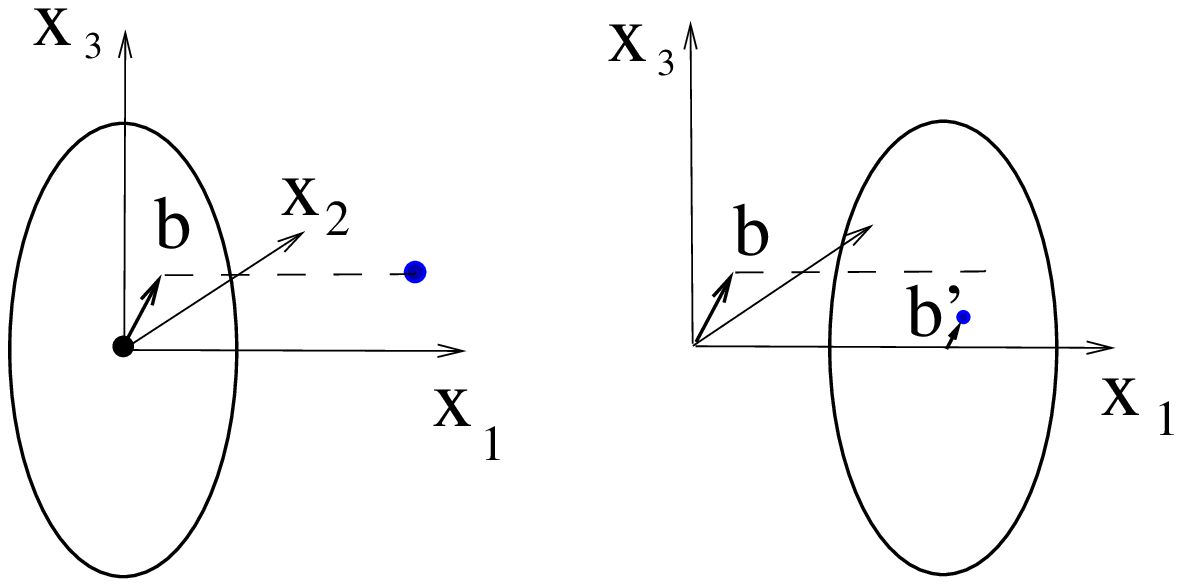}
}
\put(14,22){\includegraphics[width=0.5cm,angle=90]{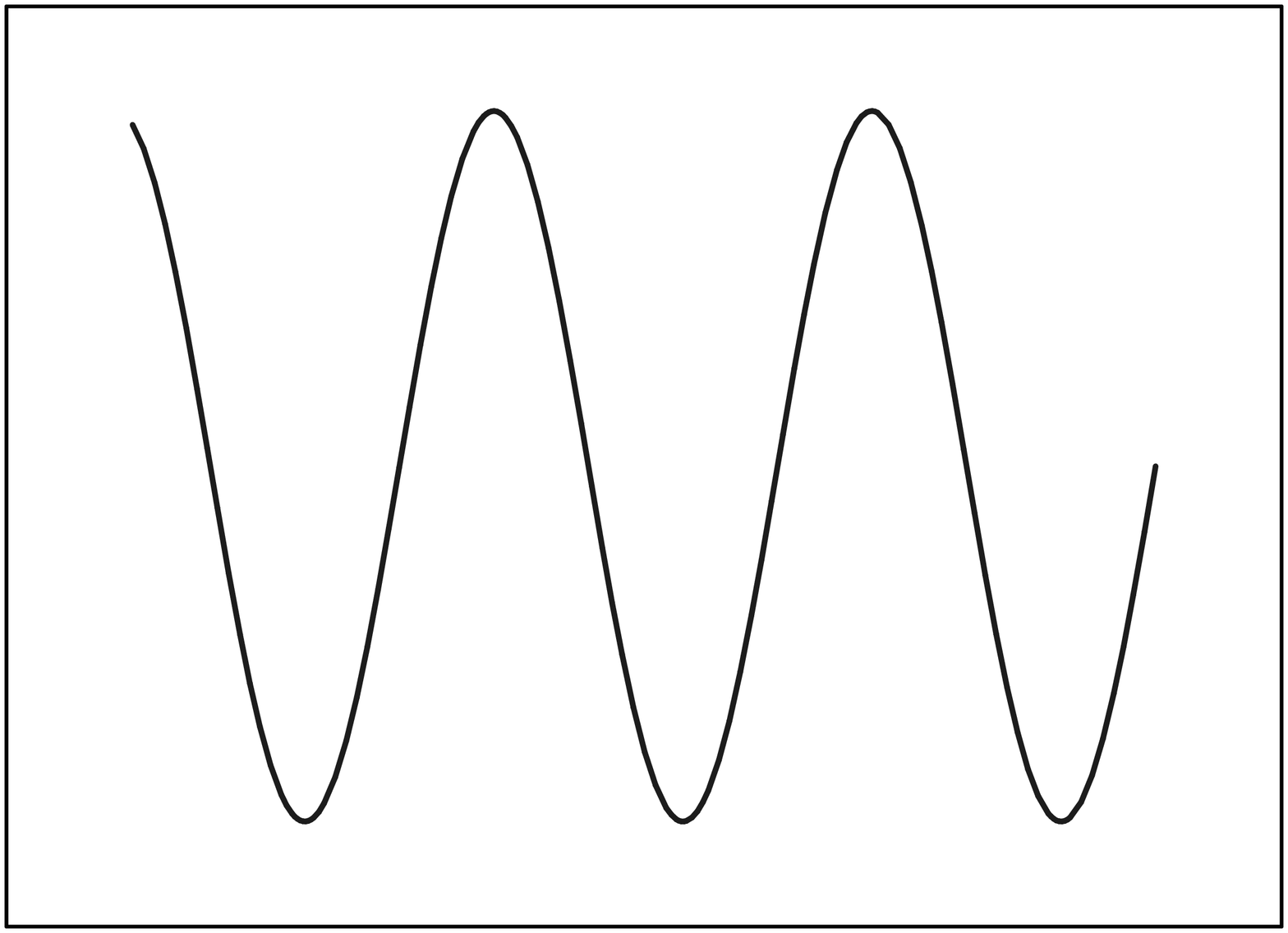}}
\put(52,18){\includegraphics[width=0.5cm,angle=90]{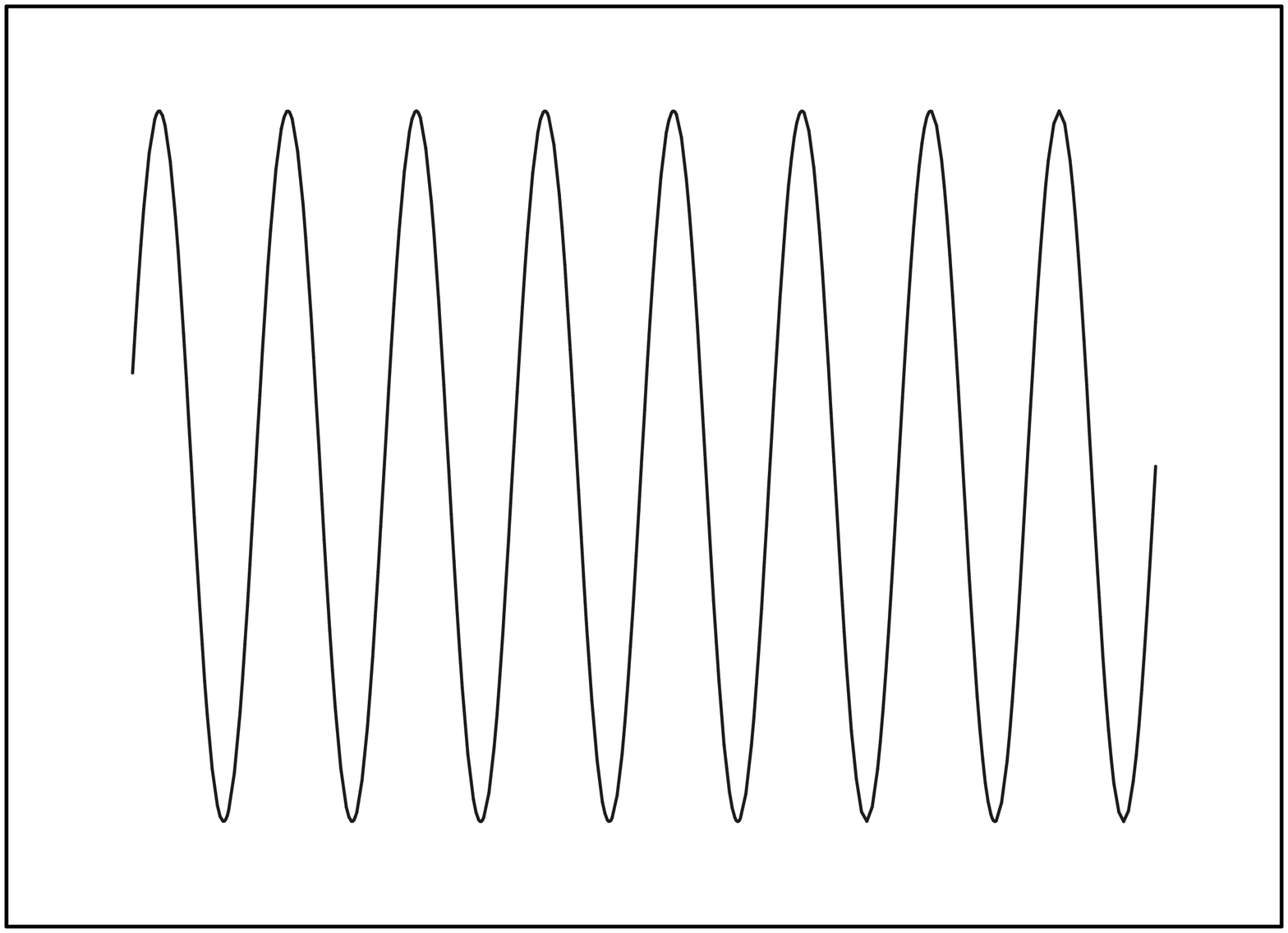}}
\put(70,0){
\includegraphics[width=3.5cm]{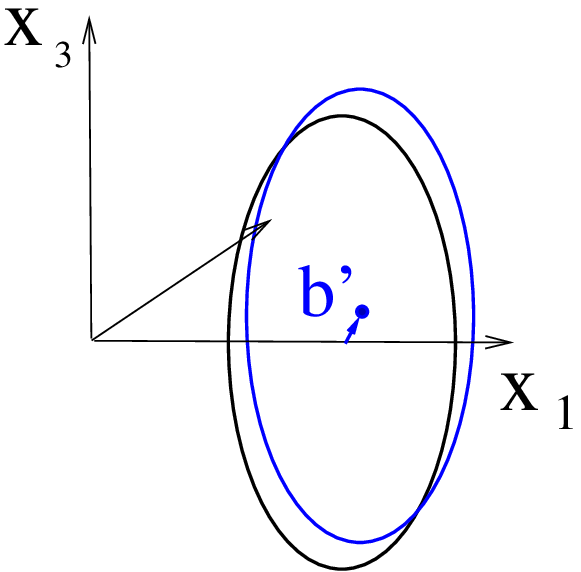}}
\end{picture}
\caption{A.  Ultra relativistic  particle (shock wave) and  a rest quantum particle,
$b$ is an impact vector.
B. Quantum particle after a collision with the ultra relativistic
particle, its impact
vector $b'$ just after collision decreases,  $|b'|<<|b|$ and its frequency increases. C.
After collision particle which was in rest after collision move with
an ultra relativistic velocity and looks as a shock wave.
 }
\label{Fig:7}
\end{figure}

It is instructive to note that the similar analyze can be done in the
in the rest frame of one of the
incident particles \cite{thooft}. This particle has large
the de Broglie wavelength and has to be treated as a quantum particle.
The gravitational field of the other, which is
rapidly moving, looks like a gravitational shock wave, see Figure \ref{Fig:7}.

Dynamics of the quantum particle
can be described by a solution of the quantum Klein-Gordon equation
in the shock wave background. This problem has been solved by 't Hooft
\cite{Hooft-eik}. Dynamics of the particle is given the eikonal approximation \cite{eikonal,BMTS}
and is defined by the geodesics behavior near  the shock
wave. The approximation is valued for a large impact parameter. The shock
wave focuses the geodesics  down to a small impact
parameter. Just in this region  we expect the BH formation (see next section)
and in this region  the eikonal approximation is not nonapplicable.
This give an explanation why a straightforward eikonal approximation
does not describe the BH production. But it is instructive to see what the
eikonal approximation can give and this is a subject of the next subsection.

The picture presented in Figure \ref{Fig:7} is idealization. More precise
approach would be
started from one moving particle with $\gamma$ rather large, but
$\gamma\neq \infty$ and other particle in the rest. It should exist a
 classical solution
that interpolates between this initial configuration and a configuration
in the later time that represents two stars which  are rather closed  and
move  slowly respect each other. One can expect to estimate  quantum fluctuations
to such classical configuration.

\subsection{BH formation and the eikonal approximation}
\begin{figure}[h!]
\centering
\setlength{\unitlength}{1mm}
\includegraphics[width=4cm]{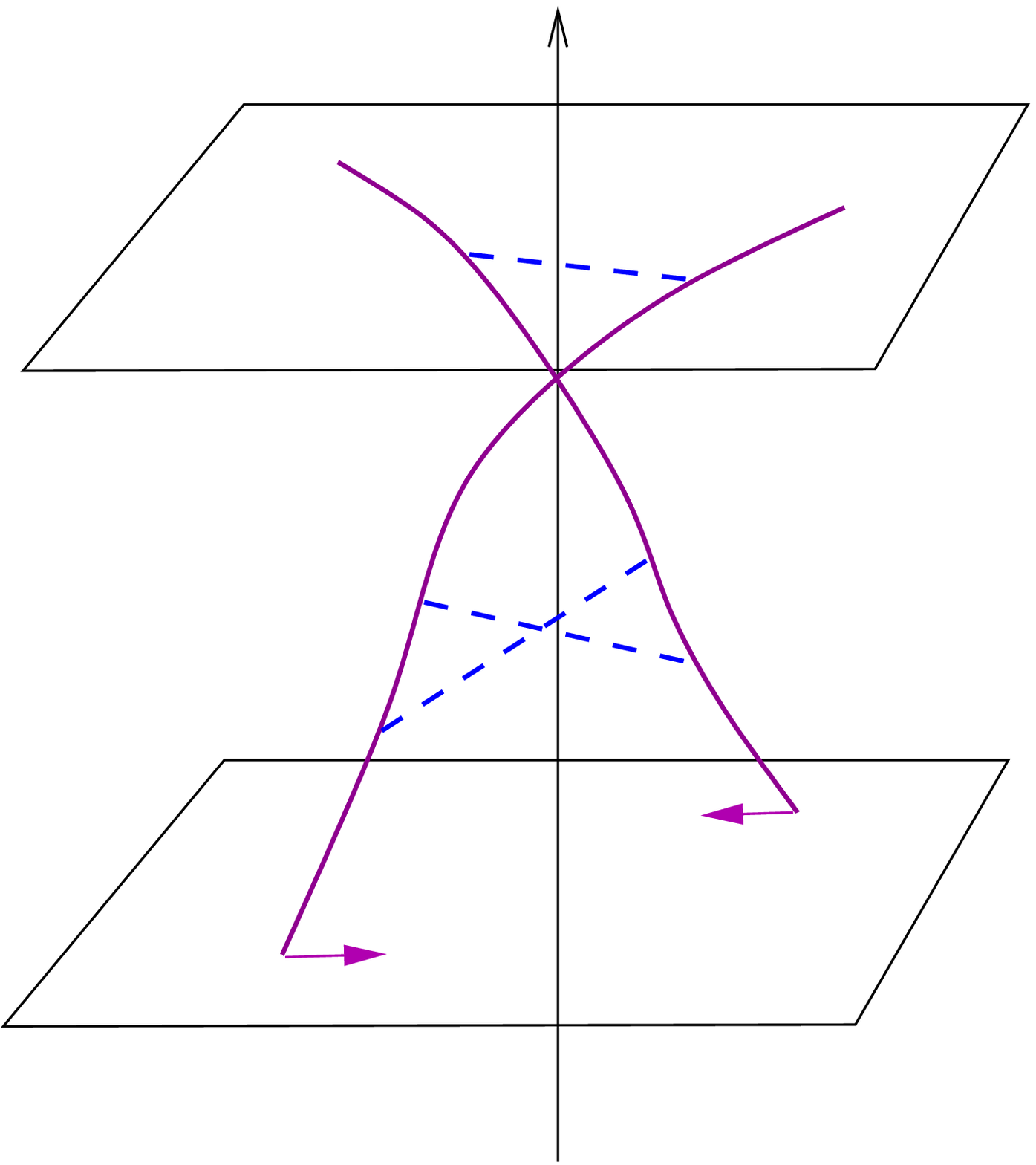}
$A\,\,\,\,\,\,\,\,\,\,$ $\,\,\,\,\,\,\,\,\,\,$
\includegraphics[width=4cm]{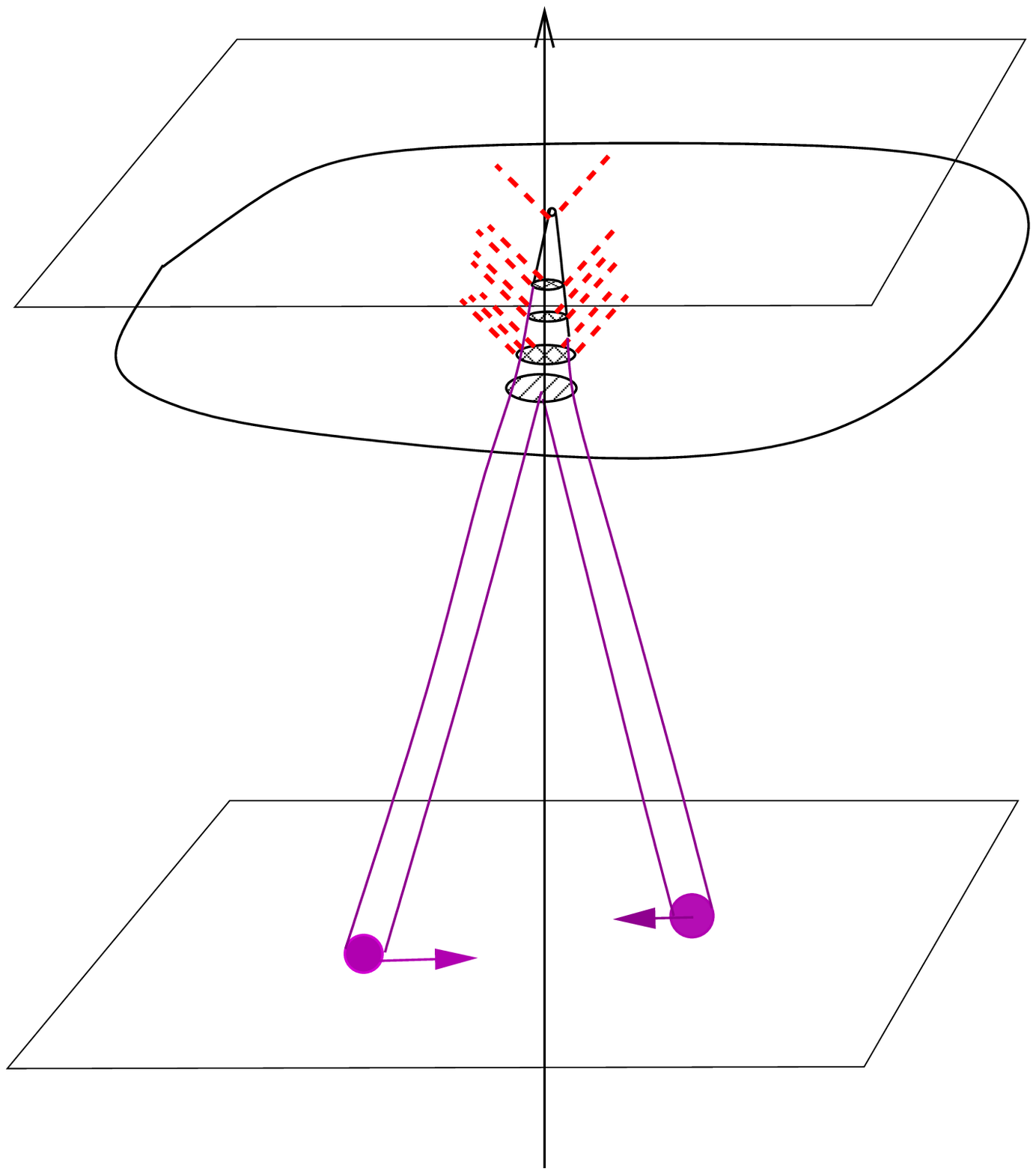}$B$
\label{Fig:sliceC}
\caption{A. Ultra relativistic colliding particles with a large impact
  parameter. Blue lines represent the graviton exchange.
  B.
  Colliding particles with a small impact
  parameter and mass/energy enough to produce BH. Red dot lines represent BH evaporation.
 }
 \end{figure}
 For a large impact parameter in the
transplanckian region
one can use the eikonal approximation
\cite{eikonal,BMTS}. Taking this  approximation  for the graviton exchange diagrams
we get \cite{Kabat,ACV,GRW},
\begin{equation}
\mathcal{A}_{eik}(\mathbf{q})=\mathcal{A}_{Born}+\mathcal{A}%
_{1-loop}+\ldots=-8 E p \int d^{2}\mathbf{b}\,e^{-i\mathbf{q}.\mathbf{b}%
}(e^{i\chi}-1)\,, \label{eq:AeikR}%
\end{equation}
with the eikonal phase $\chi$ given by the Fourier transform of the Born
amplitude in the transverse plane. The
 4-dimensional
 Born amplitude for the graviton exchange is given by
 \be
 \mathcal{A}_{Born}(\mathbf{q})=
 \frac{2 \pi G \gamma(s) }{ E p}   \frac{1}{ q_\perp^2 + \mu^2}
 \ee
here $\mu$ is IR graviton mass regularization. The corresponding eikonal phase
\cite{Kabat}, is
\be
 \chi=\frac{2 \pi G \gamma(s) }{ E p}  \int \frac{d^2 q_\perp}{(2 \pi)^2} e^{i {\bf q_\perp}\cdot {\bf x}_\perp}
  \frac{1}{q_\perp^2 + \mu^2}=
  {2 \pi G \gamma(s) \over E p}K_0(\mu b),
      \ee
where $\gamma(s)=\frac{1}{2} \left((s-2m^2)^2-2m^4\right)$,
 $K_0$ is the modified Bessel function.

 For $b\mu <<1$
 $K_0(\mu b)\sim {1\over 4 \pi}  \ln(\mu b)$ and we get
 the eikonal amplitude in term of Mandelstam variables
\be      \mathcal{A}_{eik}(\mathbf{q})
  ={ 16\pi G \gamma(s) \over -t}
      {\Gamma(1-i\alpha(s))\over\Gamma(1+i\alpha(s))}
      \left({4\mu^2\over -t}\right)^{-i\alpha(s)},\,\,\,\,
      \alpha(s) =\frac{2G\gamma(s)}{\sqrt{s ( s - 4 m^2)}}
\ee

The eikonal approximation with  a real eikonal phase
satisfies the unitarity condition
\be
\sigma_{\rm eik}= \frac{1}{16\pi^2s^2}\int d^2q_\perp
\left| {\cal A}_{\rm eik} \right|^2 = \frac{{\rm Im}{\cal A}_{\rm eik} (0)}{s}.
\label{optic-theor}
\ee
and cannot describe the black hole formation.

However, the exact knowledge of $2\to 2$
scattering amplitude can provide information about resonance states production.
Therefore,   if we know exact
$2\to 2$
scattering amplitude we expect to get an information about the BH formation
due to the unitarity condition.

There is an analogy with BH production in higher energy  and breather production
in the 2 particles scattering in
Sin-Gordon 2-dimensional model,
\be
2\,\,{\rm particles}\,\,\to\,\,\,{\rm breather}
\ee
Indeed, the classical Sin-Gordon 2-dimensional model has so-called
breather solutions with masses
\bea
\label{mn}m_{n}=\frac{16m}{\gamma}\sin \frac{n \gamma}{16},\,\,n=1,...<\frac{8\pi}{\gamma}
\eea
The exact quantum  $2\to 2$
 amplitude ${\cal A}_{\rm exact}$ for massive particles in the 2-dimensional Sin-Gordon  model
\cite{AK74} has an extra pole at
\be
M^2=4m^2_1-m^2_1(\frac{\gamma}{8})^2+...\ee
that is  nothing but the pole corresponding to the first breather.
One can see the breather contribution in the unitarity condition for
amplitude of massive particles, ${\cal A}_{\rm exact}$.

BH production in the collision of two particles can also seen as a
violation of the unitary in the $2\to 2$ elastic channel.
Indeed, let us consider a scattering
amplitude in two channels system,
\be
\mathcal{A}= \left(\begin{array}{cc}
       \mathcal{A}_{2p\to 2p} &\mathcal{A}_{2p\to BH}\\
        \mathcal{A}_{BH\to 2p} &\mathcal{A}_{BH\to BH}
        \end{array}\right)
\ee
$\mathcal{A}_{2p\to 2p}$ is the elastic scattering amplitude and
$\mathcal{A}_{2p\to BH}$ is the inelastic one.
The unitary condition means that
\be
2{\rm Im}\mathcal{A}_{2p\to 2p}=|\mathcal{A}_{2p\to 2p}|^2+|\mathcal{A}_{2p\to BH}|^2
\ee
So, if we expect $\mathcal{A}_{2p\to BH}\neq 0$ we have
a violation
of the the elastic unitarity,
\be
2{\rm Im}\mathcal{A}_{2p\to 2p}\neq |\mathcal{A}_{2p\to 2p}|^2
\ee

The simple way to break unitarity is assume that in the eikonal approximation we deal with
a complex eikonal phase. We expect the imaginary eikonal phase at small impact parameter and we write

\begin{equation}
\mathcal{A}_{eik}^{(2\to2)}(\mathbf{q})=-2s\int_{|\mathbf{b}|>b_c} d^{2}\mathbf{b}\,e^{-i\mathbf{q}.\mathbf{b}%
}(e^{i\chi}-1)-2s\int_{|\mathbf{b}|<b_c} d^{2}\mathbf{b}\,e^{-i\mathbf{q}.\mathbf{b}
}(e^{-\delta +i\chi}-1)\,, \label{eik-cut}
\end{equation}
$b_c\sim R_{S,4}$. We have
$$\sigma_{el} =\frac{1}{16\pi^2 s^2}\int \frac{d^2\mathbf{q}}{(2\pi)^2} |
\mathcal{A}_{eik}^{(2\to2)}|$$
\bea
\label{elas}
=
2\int_{|\mathbf{b}|>b_c} d^{2}\mathbf{b}\,[1-\cos\chi ]+
\int_{|\mathbf{b}|<b_c} d^{2}\mathbf{b}\,[1+e^{-2\delta}-2e^{-\delta}\cos\chi ]\eea
In accordance with the optical theorem,
$$\sigma _{total}=\frac{1}{s}{\rm Im}\mathcal{A}_{eik}^{(2\to2)}(0)$$
\bea
\label{tot}
=
2\int_{|\mathbf{b}|>b_c} d^{2}\mathbf{b}\,[1-\cos\chi ]+
2\int_{|\mathbf{b}|<b_c} d^{2}\mathbf{b}\,[1-e^{-\delta}\cos\chi ]\eea
and one can interpret the difference between (\ref{elas}) and
(\ref{tot}) as a cross section of the BH production \cite{GRW,GidPorto,LodRych}
\bea
\sigma _{BH}=\sigma _{total}-\sigma_{el}
=\int_{|\mathbf{b}|<b_c} d^{2}\mathbf{b}\,[1-e^{-2\delta}]\eea

To summarize the above discussion  we can say that to describe the BH creation
we would need to use the full classical solution describing the process, which however
is difficult to handle. From other site, the full $2\to 2$ particle amplitude
would provide  information about the BH production, but we are faraway from getting
it.
The elastic
 small-angle
scattering amplitude given by eikonalized single-graviton exchange
\cite{Hooft-eik},\cite{ACV87},\cite{Verlindes},\cite{Kabat}, valued
for large impact parameters $b\gg R_{S,4}$,
cannot describe the BH formation.
 Computing the
corrections in $b/R_{S,4}$ to the elastic scattering, one hopes to learn about
the strong inelastic dynamics at $b\sim R_{S,4}$ \cite{ACV87},\cite{Ciafaloni:2008dg},
\cite{Veneziano:2008xa}.

\newpage

\subsection{Transition amplitudes and cross section for higher dimensional gravity
and matter living on the brane }
 To consider the question of BH creation in high energy scattering in
 physical setting for low
 Planck scale we have to deal with  two   particles
confined to the  3-brane  which scatter due to the $D$-dimensional
gravitational field, $D=4+n$, $n$ is the number of large extra dimensions.
For this purpose we have to make few modifications of formula (\ref{FI-GR})
and take into account that particles interact with $D$-dimensional
gravity and with matter leaving on the brane.
\begin{itemize}
\item For simplicity we  work in $1+n$ formalism. At the initial time $t$
we deal with a slice $\Sigma$ and at the final time $t^\prime$ with a slice
$\Sigma^\prime$. The slice $\Sigma$ crosses the brane worldsheet over  3-dimensional
slice $\Xi$ and the slice $\Sigma$ crosses the brane worldsheet over  3-dimensional
slice $\Xi^\prime$.
\item Generalized coordinate include D-dimensional metric $g_{MN}$ and matter fields $\phi$
leaving on the brane $B$.
\item The state
at on a initial time is specified
by a $3+n$-metric $h_{IJ}$  on the slice $\Sigma$ and fields $\phi $ on
the slice $\Xi$ and final state by a
$3+n$-metric $h_{IJ}^\prime$ on the slice $\Sigma^\prime$ and  $\phi ^\prime$
on
the slice $\Xi^\prime$ .
\item
 The transition amplitude in this generalized coordinate representation
is given by Feynman integral, which is an extension of the formula from \cite{AVV94}
to the brane world
\be
 \nonumber
{\cal K}(h,\phi , t;h^\prime, \phi ^\prime,t^\prime)=\int e^{\frac{i}{\hbar}
S[g,\phi]}\prod
_{{\small{\small\begin{array}{ccc}&\phi |_{\tau =t}=\phi,\,\,
g |_{\tau =t}=h &\\&
\phi |_{\tau =t^\prime}=\phi ^\prime,\,\,
g |_{\tau =t^\prime}=h ^\prime &\\
\end{array}}}}~{\cal D}\phi (\tau,\vec{x}){\cal D}g(\tau,\vec{X})
\ee
where the integral is over all $4+n$-geometries which match  given values on two spacelike surfaces,
$\Sigma$ and $\Sigma^\prime$ and field configurations
which match  given values on two 3-dimensional spacelike surfaces,
$\Xi$ and $\Xi^\prime$

\item We
specify the initial configuration $h $
so that it  corresponds  to the
Minkowski brane  embedding in the bulk and matter fields $\phi  $ on
$\Xi$ ;
\begin{figure}[h!]
\centering
\setlength{\unitlength}{1.2mm}
\begin{picture}(100,30)
\put(0,0){
\includegraphics[width=3cm]{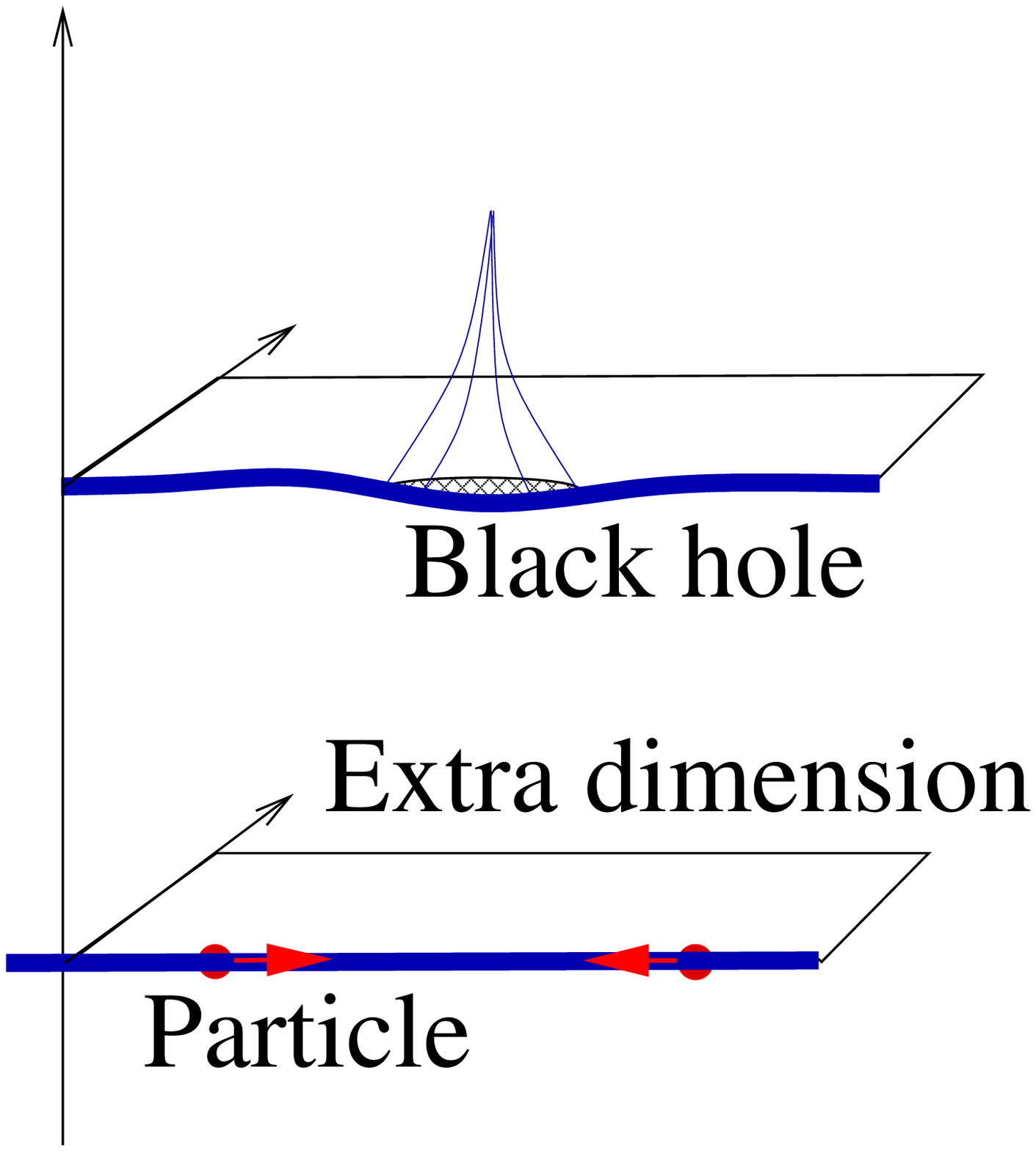}}
\put(4,19){\makebox(0,0)[lb]{$\Sigma^\prime$}}
\put(23,13){\makebox(0,0)[lb]{${\cal B}$}}
\put(32,25){\makebox(0,0)[lb]{ A slice $\Sigma ^\prime$ at $\tau =t^\prime$
 is a slice with a black hole ${\cal B}$.}}
 \put(4,12.5){\makebox(0,0)[lb]{$\Xi^\prime$}}
\put(32,20){\makebox(0,0)[lb]{Null geodesics started from the shaded region
 on $\Xi ^\prime$   }}\put(32,15){\makebox(0,0)[lb]{do not reach null infinity. }}
 \put(4,8){\makebox(0,0)[lb]{$\Sigma$}}
\put(32,7){\makebox(0,0)[lb]{A slice  $\Sigma$ at $\tau =t$ is an initial slice with
with extra }}
\put(32,2){\makebox(0,0)[lb]{dimensions and particles living on the brane $\Xi $}}
\put(32,-3){\makebox(0,0)[lb]{(blue thick line)}}
\put(3,27){\makebox(0,0)[lb]{$\tau$}}
\put(21,2){\makebox(0,0)[lb]{$\Xi$}}\end{picture}
\label{slice}
\end{figure}

$$\,$$
\item
 We specify the final configuration $h^\prime$ on $\Sigma ^\prime$ and $\phi ^\prime$
on $\Xi^\prime$
as describing black hole. ${\cal B}$ is a in D-dimensional black hole.
\end{itemize}

It is not simple to a find solution with a D-dimensional black hole and a brane.
The raison is that the usual D-dimensional black hole, say the Meyer-Perry black hole,
solves the vacuum D-dimensional
Einstein equation. But in the case of the presence of the brane the energy
momentum tensor has an extra term
providing the localization of the matter on the brane \cite{RS,kanMST,RG-VR-SS,Aref'eva:2000pe},
$T_{MN}\sim \delta(y)t_{MN}$;
see detail discussions in papers
\cite{kanWiltshire,AIV,kanMST,EHM,emp,0409099,CHR,MG1,Maeda:2009ds,RG} and recent review
\cite{Kanti}; note also the case of 1-brane
in $1+2$-space-time\cite{IA} and
the case of 2-codimensional branes, where the problem can solved for particular
examples \cite{Kaloper-2-dim}).
$$\,$$

\begin{figure}[h!]
\centering
\setlength{\unitlength}{1mm}
\begin{picture}(30,15)
\put(0,0){
\includegraphics[width=3cm]{slices-branes-1.eps}}
\put(-5,15){\makebox(0,0)[lb]{$\Sigma$}}
\put(6,4){\makebox(0,0)[lb]{$\Xi$}}
\end{picture}$A\,\,\,\,\,\,\,\,\,\,\,\,$ $\,\,\,\,\,\,\,\,\,\,\,\,$
\begin{picture}(30,15)\put(0,0){
\includegraphics[width=3cm]{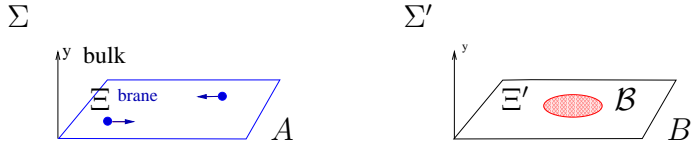}}
\put(23,4){\makebox(0,0)[lb]{${\cal B}$}}
\put(8,4){\makebox(0,0)[lb]{$\Xi^\prime$}}
\put(-5,15){\makebox(0,0)[lb]{$\Sigma^\prime$}}
\end{picture}$B$
\\
$\,$\\
\caption{Slices with brane at different times: A.  Initial slice $\Sigma$ with brane $\Xi$ and
particles on the brane and without black holes; B. Finite slice $\Sigma^\prime $ with a black hole on the brane}
\label{slice-details}
\end{figure}
$$\,$$
\begin{figure}[h!]
\centering
\setlength{\unitlength}{1mm}
\includegraphics[width=4cm]{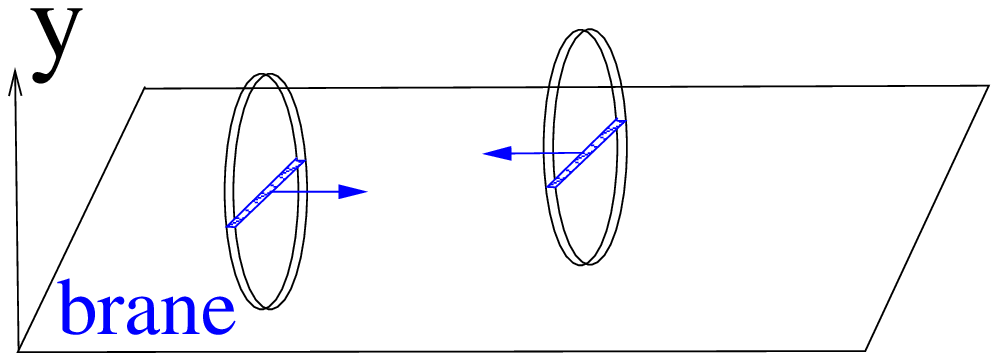}$A.\,\,\,\,\,\,\,\,\,\,\,\,\,$
\begin{picture}(30,15)\put(0,0)
{\includegraphics[width=3cm]{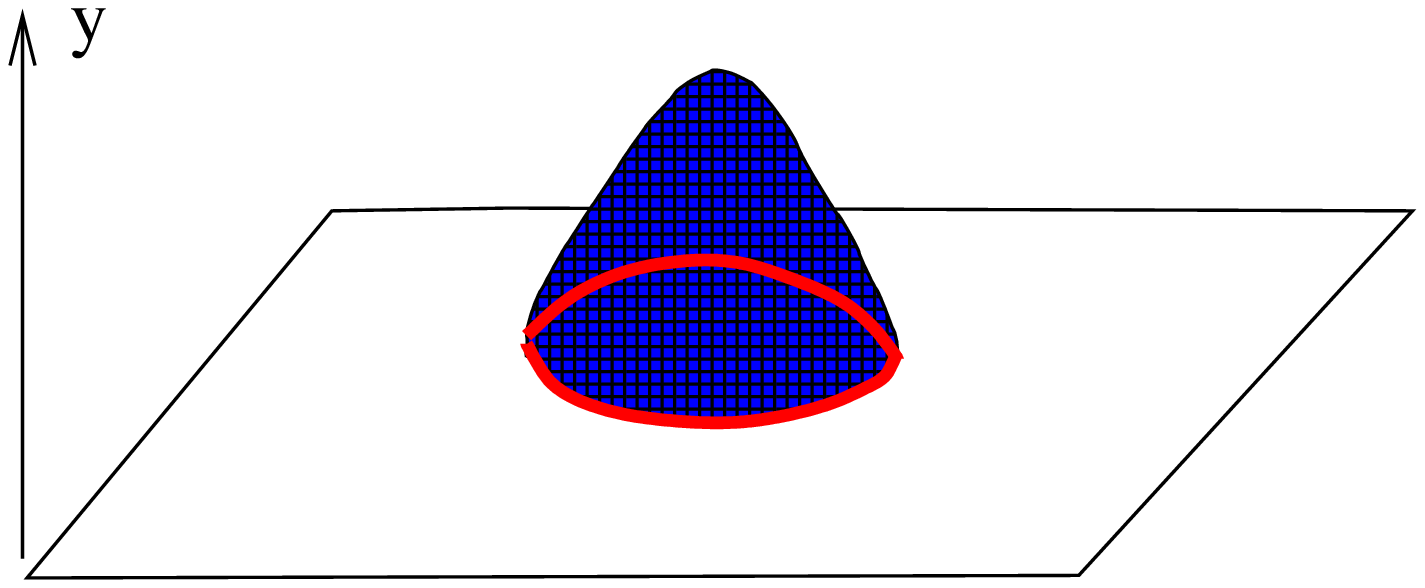}}
\put(23,4){\makebox(0,0)[lb]{${\cal B}$}}
\put(7,4){\makebox(0,0)[lb]{$\Xi^\prime$}}
\put(-5,15){\makebox(0,0)[lb]{$\Sigma^\prime$}}
\end{picture}$B$ $\,\,\,\,\,\,\,\,\,\,\,\,\,$
\begin{picture}(30,15)\put(0,-8){
\includegraphics[width=3cm]{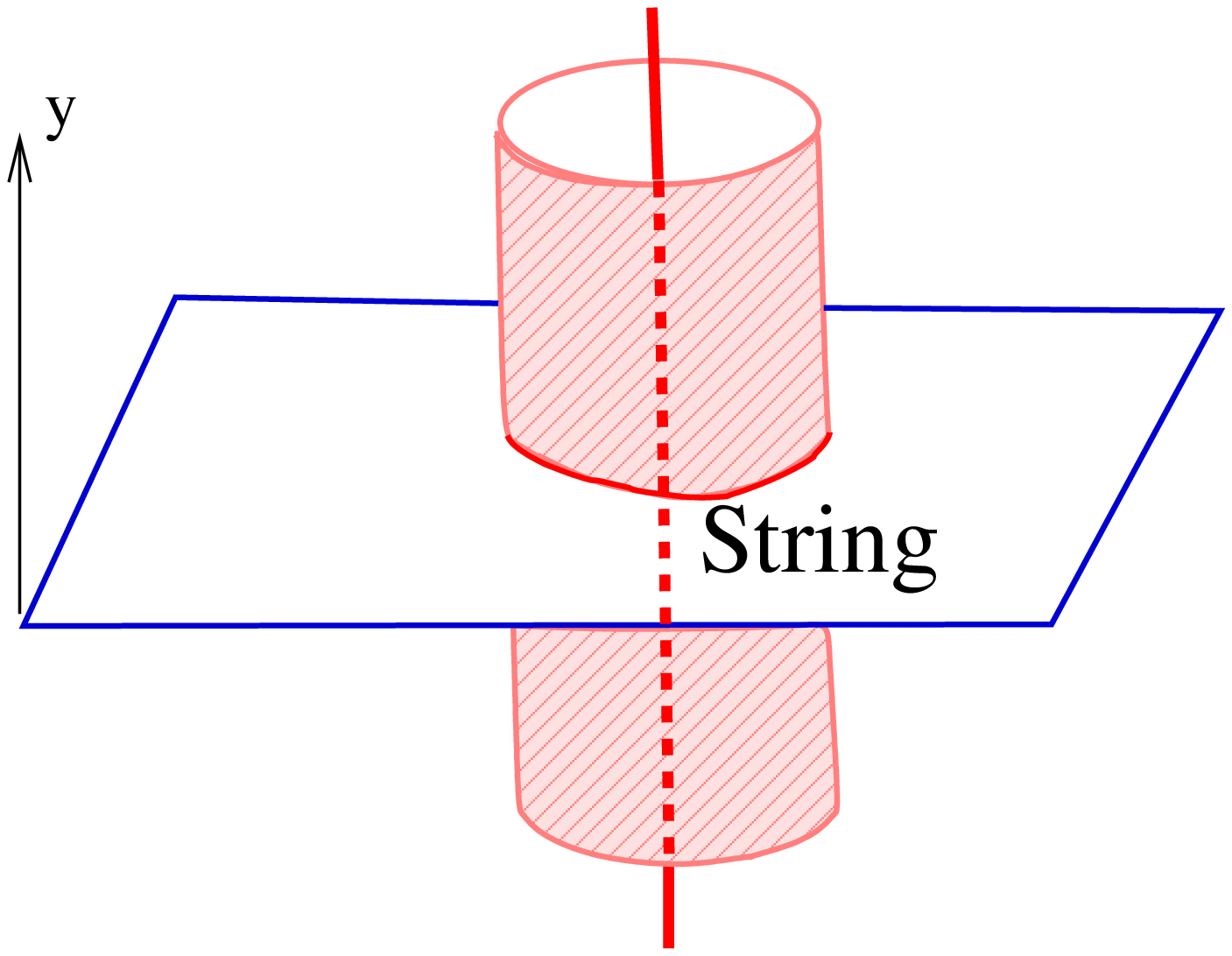}}
\put(23,4){\makebox(0,0)[lb]{${\cal B}$}}
\put(7,4){\makebox(0,0)[lb]{$\Xi^\prime$}}
\put(-5,15){\makebox(0,0)[lb]{$\Sigma^\prime$}}
\end{picture}$C$
\\
$\,$\\
\caption{A: Ultra relativistic colliding particles on the $3$-brane;
 a blue shaded region corresponds to a crossection
of the $D-2$ dimensional disk by the 3-brane; B, C: Slices with brane at final time:
B. Black hole with a source localized on a point at
the brane $\Xi^\prime$; C. Black string with a source localized on
a line along  extra dimensions }
\label{slice-details-bh-string}
\end{figure}

\newpage
\section{Black hole formation in ultra relativistic particles collision  as   a
classical gravitational collapse}
\subsection{Chargeless particles }
We consider a collision of two massive particles with
rest masses $m$ and $M$, which move towards each other with
relative  velocity $\vec {\rm v}$, and impact
parameter ${ b}$. Suppose that the particles in  the  rest frames are described by the
Schwarzschild metric with the Schwarzschild radius, $R_{S,D}(m)$  and
$R_{S,D}(M)$ given by (\ref{sh-rad}).

 For  small relative velocity   $v=|\vec {\rm v}|<<1$,
the cross section of the BH formation in the collision of these two BHs
is of the order
\be
\label{Hoop}\sigma \sim {\cal D}_{D-2} \,R^{D-2}_{S,D}(m),
\ee where  ${\cal D}_{D-2} $
is the area of $D-2$-dimensional disk given by
(\ref{D-Disk}). Here we assume $M\sim m$. Estimation (\ref{Hoop}) is based on the Thorne hoop conjecture.
This conjecture says that  an apparent horizon forms if and only if matter with
a mass $M$
gets compressed  enough such that the circumference in all directions
satisfies the condition
of ${\cal C}\lesssim 4\pi M$.

 At large relative velocities, $v \rightarrow
1$, the cross section is different and is expected to be defined not by the
rest masses but by the energy
in the c.m.f., eq. (\ref{geom-cross}).
As has been mentioned in Sect.\ref{2.5}  estimation  (\ref{geom-cross})
does not follow from the Thorne hoop conjecture.

  Below we present an  arguments in favor of  (\ref{geom-cross}) based on study of the
system of two colliding particles in the rest  frame of one of them. Our consideration
follows main steps of
Kaloper and Terning \cite{Kaloper:2007pb}. In this paper the authors considered
4-dimensional case and used the classical capture as a model of
the black hole formation.

To show (\ref{geom-cross}) following \cite{thooft,Kaloper:2007pb}
 we go to the rest frame of one of
two particles, say $M$. At large relative velocities, $v \rightarrow
1$ the gravitational field of the particle $m$ is extremely
strongly boosted in the rest frame of the particle $M$. In the infinite boost limit,
where we also take $m \rightarrow 0$ and hold $p$ fixed, the
metric reduces to an exact shock wave metric \cite{AS71,thooft}, given by
(\ref{AS}) and (\ref{D-profile}).
The metric around the
shock wave is just two  pieces of the flat space separated by the shock wave,
and test particles move freely except when they cross the shock wave front.
This picture is similar to the picture
of the electric field lines of a highly boosted charge where the lines
are compressed into the
directions transverse to its motion \cite{LLII}. Most of the scattering of a
test particle takes place while it moves through this region with a
more intense field and one can say that
the shock wave behaves as
a very thin  gravitational lens.

Before the collision the particle  M  in its own rest frame stays at the point  $X_0^1=0$,
$X_0^2=b$, $X_0^3=0$. We consider this particle  as a test particle
 in the gravitation background
(\ref{AS}) and therefore its movement after the collision is defined by the geodesics
given by
 \cite{thooft,deathpayne,higherdwaves,devega}
\bea
\label{geod-AS}
V&=&V_0+V_1U+V_f\theta(U)+V_d\theta(U)U\\
\label{tot-5}
X^i&=&X^i_{0}+X^i_{1}U+X^i_{d}\theta(U)U
\eea
with
\be
\label{Vfm}
V_f=\frac12F,\,\,\,\,\,X^i_{d}= \frac12F_{,i},\,\,\,\,
V_d=
\frac12 F_{,i}  \cdot X^i_{1}+\frac18 F^2_{,i}
\ee
 with the corresponding initial data. In $X^0,X^1$ coordinates this trajectory is
\bea
X_{(M)}^0(\tau)&=&
\tau+\frac{F}{2\sqrt{2}}\theta(\tau)+
\frac{F_{,i}^2}{16}\theta(\tau)\tau
\\
X_{(M)}^1(\tau)&=&
\frac{F}{2\sqrt{2}}\theta(\tau)
+
\frac{F_{,i}^2}{16}\theta(\tau)\tau\\
X_{(M)}^i(\tau)&=&b_{i}+\frac{1}{2\sqrt{2}}F_{,i}\theta(\tau)\tau,
\eea
here for simplicity we use $\tau=\sqrt{2}U$.

The $m$ particle in the rest frame of the M particle   moves along $U_{(m)}(\tau)=0, X^{i}_{(m)}(\tau)=0$, $i=2,3$.
If the clocks for two particles are synchronized before the collision,
i.e. $X_{(M)}^0(\tau)=X_{(m)}^0(\tau)$
for $\tau <0$, we have
\bea
X_{(m)}^0(\tau)&=&X^1_{(m)}(\tau)=\tau+\frac{F}{2\sqrt{2}}\theta(\tau)
+
\frac{F_{,i}^2}{16}\theta(\tau)\tau\\
X_i&=&0,\,\,\,\,i\geq 2\eea
The distance between the M and m particles after the collision is given by
\bea
{\cal R}(\tau)^2=(X_{(M)}^2(\tau))^2+(X_{(M)}^1(\tau)-X_{(m)}^1(\tau))^2
=b^2(1-\tau \frac{v_{f}}{b})^2+\tau^2\eea
here
\be
v_{f}=-F_{,2}
\ee

The minimal distance  is achieved  at $\tau=\tau_{min}$
\be
\label{tau-min}
\tau_{min}=\frac{b}{1+v_f^2}v_f
\ee
and is given by the formula
\be
{\cal R}_{min}(b)=\frac{b}{\sqrt{1+v_f^2}}.
\ee
In a reasonable  approximation
\be
\label{R-min-flat}
{\cal R}_{min}(b)\approx\frac{\pi M^2_{Pl,4} b^2}{p}
\ee

The relative velocity of the m and M particles after the collision is
\bea
\vec {\rm v}(\tau)=\left(\frac{\frac{d}{d\tau }(X_{(M)}^1(\tau)-X_{(m)}^1(\tau))}{\frac{d}{d\tau }(X_{(M)}^0(\tau)},
\frac{\frac{d}{d\tau }X_{(M)}^2(\tau)}{\frac{d}{d\tau }(X_{(M)}^0(\tau)},0\right)
=\left(\frac{-1
}{1+\frac{F_{,i}^2}{16}}, \frac{\frac{1}{2\sqrt{2}}F_{,2}}
{1+\frac{F_{,i}^2}{16}},0\right)
\eea
Since $F\sim p$, for large
values of $p$,  the velocity $\vec {\rm v}$ has small components.
Therefore after the
collision, in
the rest frame of the $M$ particle
 we can use the non relativistic  description and, in particular,
apply the Thorne hoop conjecture. At this point our consideration is  different
from \cite{Kaloper:2007pb}, where the capture process, related with the Laplace old idea
\cite{HawkingEllis},
 is interpreted as the BH production
 In particular,
we can say that if the minimal distance between particles after the collision
less then the  Schwarzschild  radius of the $M$ particle (in the rest frame)
 then the $m$ particle would captured by the M particle and we interpret this as a BH
formation.
The requirement that the minimal distance between particles is smaller
 than the Schwarzschild
radius of the $M$ particle,
\be
R_{S,D}(M)>{\cal R}_{min}(b),
\ee
gives a restriction on the impact parameter
\be
\label{escape-bb}
b<b_*, \,\,\,\,\,\,b_*^2=\frac{2s}{ M_{Pl,4}^4}
\ee
here we use that $s=2Mp$.

Hence for all $b$ satisfying (\ref{escape-bb}) the M particle will capture
the m particle
and  interpreting this process as the black hole formation we get a
cross section

\be
\sigma=\pi b_*^2=\frac{s}{ M^4_{Pl,4}}
\ee

This answer is in agreement with estimations of the cross section based on the trapped surface
area \cite{eardley}. These estimations are based on the area theorem which
states  that  the horizon area of the
 black hole must be greater than area of trapped surface,
giving a lower bound on the mass of the black hole.

Comparing (\ref{escape-bb}) with the restriction
of the validity of the classical description,
\be
\label{double-est}
\frac{1}{M_{Pl,4}}<b\lesssim\frac{\sqrt{s}}{M^2_{Pl,4}},
\ee
we see that the above considerations are valued only for the transplanckian energies
\be
s>M^2_{Pl,4}\ee
The right estimation in (\ref{double-est}) also means a validity of the shock wave approximation
$b<<l$ with $l$  given by (\ref{sw-validity}),
since the RHS of  (\ref{sw-validity}) is nothing but
$p/"8\pi"M_4^2$.

The above calculations are essentially more simple then the finding the trapped surface
in the case of non head-on  collision, and by this raison we call above
estimation the "express-check" of BH formation.

\subsection{Charged particles}
It is instructive to perform  the express-check  of BH formation  in the case of
charged shock waves.
These shock waves have been obtained by boosting
the Reissner-Nortstr\"om   for arbitrary D in
 spherical static (Schwarzschild) coordinates \cite{Radu}
\begin{equation}
\label{RN-lambda} ds^2=-g(R)dT^2+g(R)^{-1}dR^2+R^2d\Omega_{D-2}^2,
\end{equation}
\bea
g(R)=1-\left(\frac{R_{S,D}(m)}{R}\right)^{D-3}+\frac{Q^2}{R^{2(D-3)}},
\eea
 $R_{S,D}(m)$ is related with $m$ by (\ref{sh-rad}) and
$Q$ is related to charge $q$ as follows:
\begin{equation}
Q^2=\frac{8\pi G_Dq^2}{(D-2)(D-3)}.
\end{equation}

We note that this solution has meaning only for $R>R_{cl}$, where $R_{cl}$
is the classical radius of the charge $Q$
\be
R_{cl}=\left(\frac{\Omega _{D-2}q^2}{2(D-3)m}\right)^{1/(D-3)}\ee

Now the Schwarzschild radius $R_{S,D}(m,Q)$  depends on the value of the
charge and is the subjects of the  equation
\be
\label{root-radius}
1-\left(\frac{R_{S,D}(m)}{R_{S,D}(m,Q)}\right)^{D-3}+\frac{Q^2}{R_{S,D}(m,Q)^{2(D-3)}}
=0\ee
We note that this equation  has  solutions only for
\be
 Q^2<Q_c^2, \,\,\,\,Q_c^2=\frac{1}{4}R_{S,D}^{2(D-3)}, \,\,{\mbox or}\,\,\,|q|
 \le \frac{m}{\Omega_{D-2}} \sqrt{\frac{8\pi G_D(D-3)}{D-2}} \, .\ee
 For $Q^2<Q_c^2$ there are two solutions
 \be
R^{D-3}_{S,D \,\pm}(m,Q)=\frac{R^{D-3}_{S,D}(m)}{2}\left(1\pm\sqrt{1-
\frac{4Q^2}{R^{2(D-3)}_{S,D}(m)}}\right) \ee
The first root is very small for small $Q$
\be
R^{D-3}_{S,D \,-}(m,Q)\simeq
\frac{Q^2}{R^{D-3}_{S,D}(m)}\ee
and the second one we can consider as a small correction to the
Schwarzschild  radius of the m particle
\be
R_{S,D\,+}(m,Q)\simeq R_{S,D}(m)\left(1-\frac{1}{D-3}
\frac{Q^2}{R^{2(D-3)}_{S,D}(m)}\right)\ee

We see that for $D>3$ the
Schwarzschild radius decries  then the charge $Q^2$ increases.

The D-dimensional charged version of the Aichelburg-Sexl
metric is \cite{LS90}:
\begin{equation}
ds^2= -2dUdV+dX_i^2 +F(|X|)\delta(U)dU^2, \label{discontinuous}
\end{equation}
\begin{equation}
F(\rho)=
\left\{ \begin{array}  {cc}
-8G_4p\ln \rho-\frac{2a_4}{\rho}, & (D=4), \\
\frac{16\pi G_Dp}{(D-4)\Omega_{D-3}\rho^{D-4}} -\frac{2a_D}{(2D-7)\rho^{2D-7}}, & (D\ge 5),
\end{array}   \right.
\label{potential}
\end{equation}
where
\begin{equation}
a_D=\frac{2\pi (4\pi G_Dp_e^2)}{(D-3)} \frac{(2D-5)!!}{(2D-4)!!}%
~~(D\ge 4) \label{avalue}
\end{equation}

Therefore,
we get for $D=4$
\be
{\cal R}_{min}(b)\approx\frac{b}{v_{fQ}}=
\frac{\pi M^2_4b^2}{p}(1+\frac{\pi C M^2_4p_e^2}{p b})
\ee
and for $D>4$
\be
{\cal R}_{min}(b)\approx\frac{b}{v_{fQ}}=
\frac{\Omega_{D-3}b^{D-4}}{16\pi G_Dp }(1+\frac{32a_D\pi G_Dp}{\Omega_{D-3}b^{3D-13}})
\ee

As in the previous case, we can say that if the minimal distance between particles
after the collision
less then the  Schwarzschild  radius of the $M$ particle, we have a BH
formation.
The requirement that the minimal distance between particles is smaller
 than the Schwarzschild
radius of the $M$ particle,
\be
R_{S,D}(M,0)>{\cal R}_{min}(b,Q),
\ee
gives a restriction on the impact parameter
\be
\label{escape-b}
b<b_{*Q},
\ee
where $b_{*Q}$ is a solution of the following equation for $D=4$
\be
\label{b-star-QQ}
 R_{S,4}(m)=\frac{\pi M^2_4b_{*Q}^2}{p}(1+\frac{\pi C M^2_4p_e^2}{p b_{*Q}})\ee
and for $D>4$\be
 R_{S,D}(m)=
\frac{\Omega_{D-3}b_*^{D-4}}{16\pi G_Dp }(1+\frac{32a_D\pi G_Dp}{\Omega_{D-3}b_*^{3D-13}})
\ee

Hence for all $b$ satisfying (\ref{escape-b}) we have a BH production.
Writing  $b_{*Q}$ as
\be
b_{*Q}=b_{*}(1+Q^2x),\ee
we see that $x<0$, i.e the cross section decreases.
This result  is in an agrement with the claim that charge effects
 reduce cross sections
 of the BH production \cite{Yoshino:2006dp}.

\subsection{Charged dilaton }
\begin{figure}[h!]
    \begin{center}
\includegraphics[height=5.5cm]{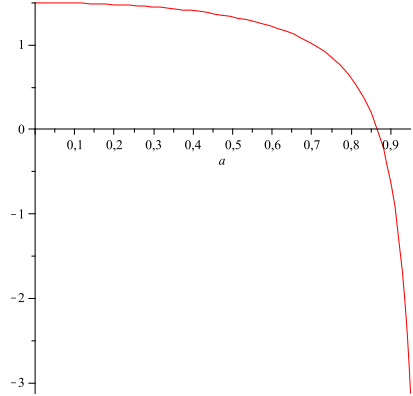}
\end{center}
\caption{Effective charge for the dilaton shock wave }
\label{Fig:a-dilaton}
\end{figure}

The four-dimensional Maxwell-dilaton part of low-energy Lagrangian obtained from
string theory is
\begin{equation}
\label{action1}
S=\frac{1}{16\pi}\int d^4x\sqrt{-g}[R-2(\nabla \phi)^2-e^{-2a\phi}F_2^2], \,\,a=1
\end{equation}
here we consider  $a$ is an arbitrary parameter. This theory has static, spherically symmetric
charged BH \cite{GG-dilaton,garf}
\be
ds^2=-A^2(r)dt^2 +A^{-2}(r)dr^2 +R^2(r)d\Omega_2^2,
\ee
where
\bea
 A^2(r)&=&\left(1-\frac{r_+}{r}\right)\left
       (1-\frac{r_-}{r}\right)^{(1-a^2)/(1+a^2)},\\
 R^2(r)&=&r^2\left(1-\frac{r_-}{r}\right)^{2a^2/(1+a^2)}.
\eea
and where $r_+, r_-$ label the two free parameters.
They are related to the physical mass and electric charge by
\be
 2M=r_+ +\frac{1-a^2}{1+a^2}r_-, \ \ \ \ \  Q^2 =\frac{r_- r_+}{1+a^2}.
\ee
After $\gamma$-boost and rescaling $M=\gamma ^{-1}p, \ \ Q^2=\gamma ^{-1}p_e^2$
 one gets \cite{dilaton-boost} the metric (\ref{AS}) with the following profile
\be
\label{prof-dilaton}
F=-\left\{ 4p \ln \rho ^2 +
  \frac{3-4a^2}{2(1-a^2)}
      \frac{\pi p^2_e}{\rho}\right\}
\ee
From (\ref{prof-dilaton}) we see that increasing/decrising of the cross section is defined by the
the sign of the the effective charge
\be
\alpha_{eff}=\frac{3-4a^2}{2(1-a^2)},
\ee
that is negative for $\sqrt{3}/2<a<1$, see Figure \ref{Fig:a-dilaton}.

Therefore for this region we have negative
values for $\alpha_{eff}$ and as result of get a catalyze of the BH production.

$$\,$$

\section{Black hole formation in ultra relativistic particle collisions in (A)dS
 as   a
classical gravitational collapse}
\subsection{Shock wave approximation of  ultra relativistic particles in (A)dS }
We consider a collision of two massive particles with
rest masses $m$ and $M$, which move towards each other in (A)dS background.
We suppose that the particles in  the  rest frames are described by the
exterior Schwarzschild (anti) de Sitter metric
\bea
\label{RN-lambda-m}
ds^2&=&-g(R)dT^2+g(R)^{-1}dR^2+R^2d\Omega_{D-2}^2,
\\
g(R)&=&1-(\frac{R_{S,D}}{R})^{D-3}-\epsilon \frac{R^2}{a^2},
\eea
where $\epsilon=+1$ for dS and  $\epsilon=-1$ for AdS,
$R_{S,D}=R_{S,D}(m)$ is related to mass $m$ as in (\ref{sh-rad}).
Now the Schwarzschild radius $R_h(m,a)$  depends on the value of the cosmological
radius and is the subject of the  equation
\be
\label{root-radius-m}
1-\left(\frac{R_{S,D}(m)}{R_{S,D}(m,a)}\right)^{D-3}-\epsilon\frac{R^2_{S,D}(m,a)}{a^2}=0\ee
We note the for $\epsilon=-1$ this equation always has a solution.
 For $a>>m$ we have
 \be
 \label{R-AdS-h}
R_{SAdS,D}(m,a)=R_{S,D}(m)\left(1-\frac{1}{D-3}
\frac{R^2_{S,D}(m)}{a^2}\right)
\ee
and we see that for $D>3$ the Schwarzschild radius increases  then the cosmological radius $a^2$
decries.

In the dS case either there are two solutions of (\ref{root-radius}) or there is no solutions at all.
The minimal cosmological radius $a^2$ below which there are no solution is
\be
a_c^2=c_DR^2_{S,D}(m),
\ee
where
$
c_D=(D-1)/(D-3)\left((D-1)/2\right)^{1/(D-3)}
$
and in particular, $c_4=\frac{27}{4}$.

For $a>>a_c$ the minimal root of equation (\ref{root-radius}) with $\epsilon =1$ is
 \be
 \label{R-dS-h}
R_{SdS,D}(m,a)=R_{S,D}(m)\left(1+\frac{1}{D-3}
\frac{R^2_{S,D}(m)}{a^2}\right)
\ee

 When these two particles
 move  towards each other in the (A)dS space-time with small cosmological constants
 and with a
 small velocity,
the cross section of the BH formation in this collision
is of order $\sigma \sim \pi R^{D-2}_{S(A)dS,D}(m)$ and as in the flat case are
 just neglible numbers.

We can increase the cross section of the BH formation by increasing up to
ultra-relativistic the relative velocity of the colliding particles.
To describe this collision in the (A)dS backgrounds in is convenient
to describe in In terms of the dependent plane coordinates,
$U,V, X^2...X^{D-1},X^D$, $\vec X=(X^2,\dots,X^{D-1})$, satisfying
\be
 -2UV+X_2^2+...+X_{D-1}^2+\epsilon X_{D}^2=\epsilon a^2
 \ee
As in the flat case to consider the collision of two particles
with  large relative velocities, $v \rightarrow
1$,  we consider this process from the rest frame of one of these two particles,
say the $M$ particle.  The
 gravitational field of the $m$ particle is extremely strongly boosted.
 In this limit, we can approximate the field by the linearized Schwarzschild (A)dS
 metric, boosted to a very large velocity.
 In the infinity boost limit, where we also take  $m \rightarrow 0$ and hold $p$ fixed,
 the Schwarzschild (A)dS
 metric reduces to a shock metric
 \cite{tanaka,emp,Sfetsos,GrifPod,Podolsky:2002nn,0606126}.
 In terms of the dependent plane coordinates,
 the line element of the shock wave
space-time is
\be
\label{SW}
ds^2=-2dU\,dV+d\vec X^2+\epsilon dX_{D}^2+F(\vec X)\delta(U)dU^2.
\ee
The shock wave shape function $F$ is a fundamental solution of the equation
\be
\left(\triangle_{D-2}+\epsilon\frac{D-2}{a^2}\right)F=
-16\sqrt{2}\pi G_D\bar p\delta(\vec n,\vec n_0),
\label{ee1}
\ee
where $\triangle_{D-2}$ is the Laplace--Beltrami operator on a
$(D{-}2)$-dimensional sphere $\mathbb{S}^{D-2}$ in the $dS$ case
or on a $(D{-}2)$-dimensional hyperboloid  $\mathbb{S}^{D-2}$ in the $AdS$ case,
$\vec n=\vec x/|\vec x|$, $\vec n_0$ is the location of the particle
on the sphere or the hyperboloid,  $\bar p$ is the energy of the shock wave, and $G_D$ is the
$D$-dimensional gravitational constant.

For $D=4$, $\epsilon=1$ we deal with
\be\label{eq:F4}
F^{dS}_4=p\left(-1+\frac{Z_4^0}{2a}\ln \frac{a+Z_4^0}{a-Z_4^0}\right)\ee
where $p$ is related with $p_{ABG}$ from \cite{ABG},
$
p= 8\sqrt2 p_{ABG}=8\sqrt2\bar p_{ABG} G_4$.

For $D=5$, $\epsilon=1$
 \bea F^{dS}_5(\xi)&=&\frac{3\sqrt{2}\pi p_5}{a}\frac{2\xi^2-1}{\sqrt{1-\xi^2}}\eea
   where $\xi=Z_5/a$ and $p_5=\bar{p}G_5$.

For $D=4$, $\epsilon=-1$ we deal with the shock wave in the  AdS space-time
\be
F^{AdS}_4=p\left(-1+\frac{Z_4^0}{2a}\ln \frac{a+Z_4^0}{-a+Z_4^0}\right)\ee
A relation with  notations in \cite{IA-AB-TMP} is
$p= 4\sqrt2 p_{AB}=4\sqrt2\bar p_{AB} G_4$.

For $D=5$, $\epsilon=-1$
\be
F^{AdS}_{5}(\bar{M},Z_5)
= -\frac{p}{a}\,\left(\frac{1
-\frac{2Z_5^2}{a^2}}{ \sqrt{\frac{Z_5^2}{a^2}-1}} +
\frac{2Z_5}{a}\right)
\ee
and $
p= 3\pi \bar{M}_{ABJ}$, \cite{ABJ}.

\subsection{Geometrical Picture of BH production in Particle Collision in
(A)dS}
The metric around the
shock wave (\ref{SW}) is just two  pieces of (A)dS space separated by the shock wave,
and test particles move along timelike geodesics of (A)dS space.
From the point of view  of the flat dependent coordinates the particles
move freely except when they cross the shock wave front.
This picture is similar to the picture in the flat case and the shock wave behaves as
a very thin  gravitational lens.
The amount of bending of the geodesics crossing the shock   is  proportional to
$p$.
Thus for $\gamma \to \infty $ we can get scattering at almost
right angle in the $U-X^1$ plane, and a particle with a large
impact parameter   after crossing the shock wave  passes very
close to the path of the boosted particle.
 If the scattered
particles end up within a distance smaller than the horizon of the
target particle then we get a black hole formation.

Let us show this explicitly.
We consider the geodesics starting from the point $Z_4^0\neq 0$,
$Z_2^0\neq 0$ $Z_3^0=0,$
and use the parametrization
\bea
AdS:&\,& Z_4^0=a\cosh\vartheta,\,\,\,\,Z_2^0=a\sinh\vartheta, \\
dS:&\,&Z_4^0=a\cos\vartheta,\,\,\,\,\,\,\,Z_2^0=a\sin\vartheta.
\eea

For these initial data for timelike geodesics we have \cite{Podolsky:2001vu,ABG}:
\bea
V( U) & = &  V^0\,S(U)\ +\ V_1\, U
  \ +\ B\,\Theta( U)\,S(U) +\ C\,\Theta( U)\, U\ , \label{geod-V-ads-m}\\
   Z_2( U) & = & Z_2^0\,S(U)
   \ +\ A_2\,\Theta( U)\, U \ , \label{geod-Z2-ads-m} \\
   Z_3( U) & = &
   0, \label{geod-Z3-ads-m} \\
   Z_4( U) & = & Z_4^0\,S(U)
   \ +\ A_4\,\Theta( U)\, U \ , \label{geod-Z4-ads-m}
\eea
where
\bea
 A_i& =&
  - \frac{1}{6}\Lambda\,Z_i^0G(0), \,\,\,\,i=2,3, \,\,\,
 A_4 =   \frac{1}{2}\left[\epsilon F_{,4}(0)
  -\frac{1}{3}\Lambda\,Z_4^0G(0)\right]\ , \nonumber \\
 \label{sol_BCA}  B &=& \frac{1}{2}F(0)\ ,
 C = \frac{1}{8}\left[\epsilon F_{,4}^2(0)+
   \frac{1}{3}\Lambda\,F^2(0)
  -\frac{1}{3}\Lambda\left(Z_4^0\,F_{,4}(0)\right)^2\right]\ ,
\eea
and
\be
F(0)\equiv F(Z^0_4),\,\,\,\,G=Z_p\,F_{,p}-F
\ee
\be
\label{factor-U}
S(U)=\sqrt{1+{\textstyle{1\over3}}\Lambda\,{(\dot{ U}^0)}
   ^{-2}\, U^{\,2}},\,\,\,\,\,\, \,\,\,\,\,\, V_1=\dot V^0/\dot U^0\ee

From the above formula we can estimate the minimal distance between the particles.

In term of $X^0,X^1$ coordinate for the $M$ particle we have the following coordinates

\bea
Z_{(M)}^0&=&\frac{V+U}{\sqrt{2}}=\frac{V^0}{\sqrt{2}}\,S(U)
 + \frac{V_1+1}{\sqrt{2}}\, U
   + \Theta( U)\frac{B\,S(U) + C U}{\sqrt{2}}
\label{target-Z0}\\
Z_{(M)}^1&=&\frac{V-U}{\sqrt{2}}=\frac{V^0}{\sqrt{2}}\,S(U)
 + \frac{V_1-1}{\sqrt{2}}\, U
   + \Theta( U)\frac{B\,S(U) + C U}{\sqrt{2}}
  \label{target-Z1}\\
 Z_{(M)}^2( U) & = & Z_2^0\,S(U)
   \ +\ A_2\,\Theta( U)\, U \ , \label{target-Z2}
     \\
   Z_{(M)}^4( U) & = & Z_4^0\,S(U)
   \ +\ A_4\,\Theta( U)\, U \ , \label{target-Z4} \\
    Z_{(M)}^3( U) & = &0
  \label{target-Z3}
\eea
Since in the coordinate system we deal with  the times are synchronized
we suppose that
\be
Z_{(m)}^0=Z_{(M)}^0
\ee
For the $m$ particle we have
\bea
Z_{(m)}^0&=&\frac{V+U}{\sqrt{2}}=\frac{V^0}{\sqrt{2}}\,S(U)
 + \frac{V_1+1}{\sqrt{2}}\, U
   + \Theta( U)\frac{B\,S(U) + C U}{\sqrt{2}}
\label{shock-Z0}\\
Z_{(m)}^1&=&Z_{(m)}^0, \\
Z_{(m)}^4&=&a\\
Z_{(m)}^i&=&0, i=2,3
\eea
The distance is
\be
{\cal R}^2_{(A)dS}((M),(m),U)
=(Z_{(m)}^1-Z_{(M)}^1)^2
+(Z_{(m)}^2-Z_{(M)}^2)^2+(Z_{(m)}^3-Z_{(M)}^3)^2\ee
Taking into account   that $Z_{(m)}^1=Z_{(m)}^0$ and $Z_{(m)}^0=Z_{(M)}^0$
we get
\be
{\cal R}^2_{(A)dS}((M),(m),U)=2U^2
+Z_2^{0^2}\left(\sqrt{1+\frac{\epsilon}{a^2} U^2}+\frac{A_2}{Z_2^0}U\right)^2
\ee
Comparing  this answer with the similar answer in the flat background,
 we see that their are similar, except the factor $S(U)$ given by
 (\ref{factor-U}) instead of the unit.

For $dS_4$ case
 \be
{\cal R}^2_{dS}(\tau)= \tau^2
+b^2\left(\sqrt{1+\frac{\tau^2}{2a^2} }-\frac{B_2}{b}\tau\right)^2,
\ee
where
\be
\label{B2}
B_2=\frac{p }{2b\sqrt{2}}, \,\,\,\,\,b=a\sin\vartheta,
\ee
$R^2_{dS}(\tau)$ as a function of $\tau$ has one minimum at the real point
$\tau_{min}$.

The expression for $\tau_{min}$ can be expanded on $1/a^2$
\be
\tau_{min} \approx b\,\frac{B_2}{B_2^2+1}+b_0\,\frac{b_0^2}{4a^2}\,
\frac{B_2(B_2^2-2)}{(B_2^2+1)^3}+{\cal O}(1/a^4)
\ee
(here we assume that $b>0$ ). Note that in the leading order
\be
\tau_{min}\approx b\,\frac{B_2}{B_2^2+1} {\cal O}(1/a^2)
\ee
that is in agreement with (\ref{tau-min})

Under assumption $B_2>>1$ we get
\be
\label{tau-min-dS}
\tau_{min,dS_4}=
\frac{2\sqrt{2}b^2}{p}+\frac{2b^5}{a^2p^2 }
\ee
and, therefore
\be
\label{R-min-dS}
{\cal R}_{dS_4,min}(b)\approx
\frac{2\sqrt{2}b^2}{p}
(1+\frac{2b^4}{a^2p^2})
\ee

For $AdS_4$ case
\be
{\cal R}^2_{AdS_4}(\tau)= \tau^2
+b^2\left(\sqrt{1-\frac{\tau^2}{2a^2} }-\frac{B_2}{b}\tau\right)^2
\ee
and
\be \label{B2-AdS}
B_2=\frac{p }{2b\sqrt{2}},\,\,\,\,\,b=a\sinh\vartheta,
\ee
As for $dS$ case, under assumption $B_2>>1$ we get
\be
\label{tau-min-AdS}
\tau_{min,AdS_4}=
\frac{2\sqrt{2}b^2}{p}-\frac{2b^5}{a^2p^2 }
\ee
and, therefore
\be
\label{R-min-AdS}
{\cal R}_{AdS_4,min}(b)\approx
\frac{2\sqrt{2}b^2}{p}
(1-\frac{2b^4}{a^2p^2})
\ee
Comparing (\ref{R-min-flat}), (\ref{R-min-AdS}) and (\ref{R-min-dS})
 we see that for the same values of $b^2/p$ the value of
 $R_{min}(b^2/p)$ is minimal for the AdS case.

 If the scattered
particles end up within a distance smaller than the horizon of the
target particle then we certainly expect a black hole to form.
Therefore, to estimate the geometrical cross section we find all impact parameters for which
takes place the condition
\be
R_{S,D}(M)>R_{min}(b),
\ee
where $R_{S,D}(M)$ is the Schwarzschild radius (without the cosmological constant).
Taking into account estimations of the minimal distances (\ref{R-min-dS}) and (\ref{R-min-AdS})
we get
\be
\label{b-rest}
R_{S,D}(M)>\frac{2\sqrt{2}b^2}{p}
(1+\epsilon\frac{2b^4}{a^2p^2})
\ee

At the limit $a\to \infty$ we have
\be
2MG_4>\frac{2\sqrt{2}b^2}{p},
\ee
and this relation holds for all $b^2<b^2_0=MG_4p/\sqrt{2}$,

We see that for the dS case the LHS of (\ref{b-rest}) as a function of $b$ is an
increasing function. In the case of the AdS it is increasing up to $b_{max}=(a^2p^2/6)^{1/4}$.
Therefore, restriction  (\ref{b-rest}) holds for all
$b<b_*$, where
the critical value $b_*$ satisfies  the relation
\be
R_{S,4}(M)=\frac{2\sqrt{2}b_*^2}{p}
(1+\epsilon\frac{2b_*^4}{a^2p^2})
\ee
and in the AdS case we also assume that $b_*<b_{max}$.

For the $AdS_5$  case
\be
{\cal R}_{min}\approx\frac{b}{B_2}
(1-\frac{b^2}{4a^2B^2_2})
\ee
Taking into account that in 5-dim case
\be
B_2=\frac{p}{2\sqrt{2}b^2},
\ee
we get
\be
{\cal R}_{min,5}\approx\frac{b}{B_2}
(1-\frac{b^2}{4a^2B^2_2})=\frac{2\sqrt{2}b^3}{p}
(1-\frac{2b^6}{a^2 p^2})
\ee
and from this formula we see that cross section in AdS case more then the
crossection in the flat case for the same value of $p$. Therefore the negative cosmological
 constant catalyzes the BH production.

\newpage

\section{Conclusion}
We rederive  the classical geometrical cross section of  the BH
production reconsidering the process of two transplanckian particles
collision  in the rest frame of one of incident particles.
This consideration permits to use the standard Thorne's hoop
conjecture for a  matter compressed into a region to prove
a variant
of the Thorne's hoop
conjecture dealing with a total amount of  compressed energy
in the case of colliding particles.

We show that the process of BH formation is catalyzed by the negative cosmological
constant and by a special scalar matter.
In opposite, it is relaxed by the positive cosmological constant and at a  critical
value just turns off.
Also we note that
the cross section is sensible  to the compactification
of extra dimensions and particular brane models and this will be studied in
separated paper in  details.

\section*{Acknowledgments}

The author is grateful to I.V.Volovich for fruitful discussions.
This work  is supported in part by RFBR grants 08-01-00798 and 09-01-12179-ofi$_m$
and  by state contract of Russian Federal
Agency for Science and Innovations 02.740.11.5057.

\newpage
\appendix
\section{BH as an initial/final data
}

Let $({\cal M},g)$ is the space-time with a metric, ${\cal M}$ is a manifold.

Black holes  are conventionally defined in
asymptotically flat space-times by the
existence of an event horizon $H$.

The horizon $H$ is the boundary
 of the causal past
of future null infinity, i.e. it is the boundary of the
set of events
in space-time from which one can escape to infinity in the future
direction.

To be more precise we need few definitions\cite{HawkingEllis,Ward}.

All
notions  work for BH production in the particle collision as well as for gravitational
collapse.

\subsection{(Weakly) Asymptotically
Simple  Space-time}
A (oriented in time and space) space-time $({\cal M}, g_{\mu \nu})$
is {\it asymptotically simple} (\cite{HawkingEllis}, p.246) if there
exists a smooth manifold $\tilde {{\cal M}}$
with metric $\tilde {g}_{\mu\nu}$,
boundary ${\cal I}$, and a scalar
function  $\Omega $ regular everywhere on
$\tilde {{\cal M}}$ such that
\begin{itemize}
\item $~~\tilde {{\cal M}}-{\cal I}$ is conformal
to ${\cal M}$ with  $\tilde {g}_{\mu\nu}=
\Omega ^{2} g_{\mu\nu}$,
\item $~\Omega >0$ in $\tilde {{\cal M}}-{\cal I}$
and     $\Omega =0$ on ${\cal I}$
with $\nabla _{\mu }\Omega \neq 0$ on ${\cal I}$,
\item Every null geodesic
on $\tilde {{\cal M}}$ contains, if maximally extended,
two end points on  ${\cal I}$.
\end{itemize}

 ${\cal I}$
consists of two disjoint pieces ${\cal I}^{+}$ (future null infinity)
and ${\cal I}^{-}$ (past null infinity)
each topologically $R$x$S^{2}$,
\be
{\cal I}={\cal I}^{+}\cup{\cal I}^{-}\ee

One can symbolically write
\be
\tilde {{\cal M}}={\cal M}\cup \partial {\cal M},\ee
where $ \partial {\cal M}={\cal I}$.

If $M$ satisfies the Einstein vacuum equations near ${\cal I}$
then ${\cal I}$ is null.

A space-time ${\cal M}$ is {\it weakly asymptotically simple}
 if there exists
an asymptotically simple $M_{0}$ with corresponding $\tilde{{\cal M}}_{0}$
such that for some open subset K of  $\tilde{{\cal M}}_{0}$
including ${\cal I}$,
the region ${\cal M}_{0}\cup K$ is isometric to an open
 subset of ${\cal M}$.  This
allows ${\cal M}$ to have more infinities than just ${\cal I}$.

\subsection{Asymptotically
Flat  Space-time}
A space-time  is {\it asymptotically flat} if it is
{\it weakly asymptotically simple and empty}, that is, near future and
past null infinities it has a conformal structure like that of Minkowski
space-time.
\subsection{Global (Partial) Cauchy surface}
Let ${\cal S}$ be a space-like hypersurface. The
future(past) domain of dependence of ${\cal S}$), denoted $D^+({\cal S})$($D^-({\cal S})$), is
defined by \cite{HawkingEllis}
\be
D^{\pm}({\cal S})=\left\{p\in {\cal M}|\mbox {Every past(future) inextensible causal curve
 through } p \,\,\mbox {intersect } {\cal S} \right\}\ee

The full domain of dependence of ${\cal S}$ is defined as
\be
D({\cal S})=D^+({\cal S}) \cup D^-({\cal S})
\ee

The set ${\cal S}$ for which $D({\cal S})={\cal M}$ is called a Cauchy surface,
or a {\it global Cauchy surface}.
Causal curves cross the  global Cauchy surface just one time.

\begin{figure}[h!]
\centering
\includegraphics[width=5cm]{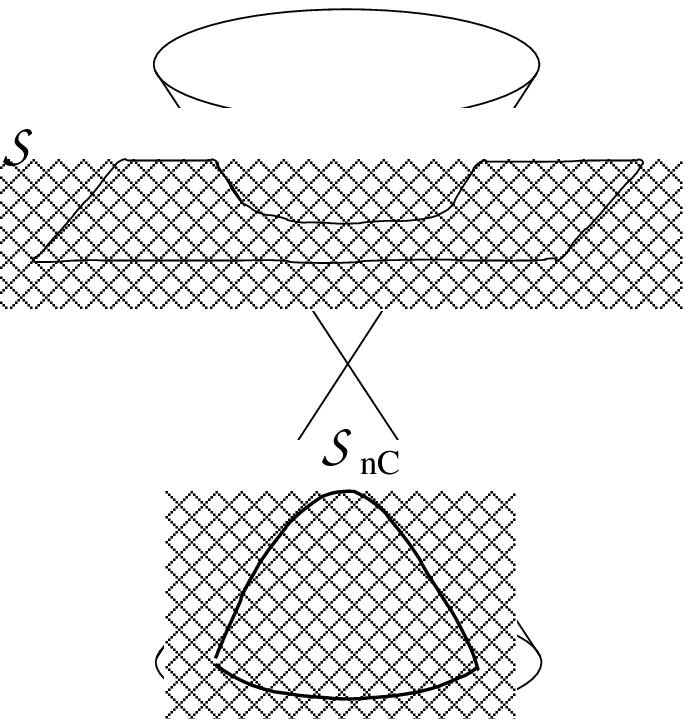}
\caption{A. The Minkowski space time $M^4$ with the Cauchy surface ${\cal S}$
and a surface ${\cal S}_{nC}$ which is not the Cauchy surface, \cite{HawkingEllis},
Fig.13}\label{Fig:D1}
\end{figure}

The     space-like hypersurface ${\cal S}$ is called the {\it partial Cauchy surface}
 if
non one causal curve crosses  it more then one time \cite{HawkingEllis},
see Figure \ref{Fig:D1}.
$$\,$$

\subsection{Causal Future $J^+(p)$ and Causal Past $J^-(p)$
of the  Point }
A causal future $J^+(p)$( past $J^-(p)$) of the point $p$ is defined as
\bea
 J^{\pm}(p)&=&\{q\in {\cal M},\mbox{\,such that there
is future (past) oriented}\\
&\,&\,\,\,\,\mbox{ causal curve} \,\,\,\gamma(\tau), \mbox{\,\, so that}\,\, \gamma(0)=p,\gamma(1)=q\}
\nonumber
\eea

A causal future $J^+(\Sigma)$ of the surface $\Sigma$ is defined as
\bea
 J^-( \Sigma)&=&\{q\in {\cal M},\mbox{\,if  there
is future oriented}\\
&\,&\,\,\,\,\mbox{ causal curve} \,\,\,\gamma(\tau), \mbox{\,\, so that}\,\,
\gamma(0)=p\in \Sigma,\,\,\gamma(1)=q\}
\nonumber
\eea
In particular,
\bea
 J^-( {\cal I}^+)&=&\{q\in {\cal M},\mbox{\,if  there
is future oriented}\\
&\,&\,\,\,\,\mbox{ causal curve}\,\,\, \gamma(\tau), \mbox{\,\, so that}\,\,
\gamma(0)=p\in {\cal I}^+,\gamma(1)=q\}
\nonumber
\eea

\subsection{Black Holes in Asymptotically
Flat  Space-time}
Black holes  are conventionally defined in
asymptotically flat space-times by the
existence of an event horizon $H$.
The horizon $H$ is the boundary $\dot {J}^{-}({\cal I}^{+})$
 of the causal past $J^{-}({\cal I}^{+})$
of future null infinity ${\cal I}^{+}$, i.e. it is the boundary of the
set of events
in space-time from which one can escape to infinity in the future
direction.

The {\it black hole region } $B$ is $$B=M-J^{-}({\cal I}^{+})$$
and {\it the event horizon}
$$H=\dot {J}^{-}({\cal T}^{+}).$$

\subsection{Future (strongly) Asymptotically
Predictable  Space-time}
A space-time is {\it future asymptotically
predictable} if there is a surface ${\cal S}$ in spacetime that serves
as a Cauchy surface for a region
extending to future null infinity\footnote{This notion gives a formulation of
Penrose's cosmic censorship
conjecture.}.

This means that there are no "naked
singularities"
(a singularity that can be seen from infinity)
to the future of the surface ${\cal S}$.

Let $\Sigma $ be  a partial Cauchy surface in a weakly
 asymptotically simple
and empty space-time $(M,g)$. The space-time $(M,g)$ is (future) {
\it asymptotically
predictable from} $\Sigma$ if ${\cal I}^{+}$ is contained
in the closure
of $D^{+}(\Sigma)$ in $\tilde{M}_{0}$.

If, also, $J^{+}(\Sigma) \cap \bar{J^{-}}({\cal I}^{+},\bar{M})$
is contained in $D^{+}(\Sigma)$ then the space-time $(M,g)$
is called {\it strongly   asymptotically predictable}  from $\Sigma$.
In such a space there exist a family $\Sigma (\tau)$, $0<\tau<\infty$,
of spacelike surfaces homeomorphic to $\Sigma$ which cover
$D^{+}(\Sigma)-\Sigma$ and intersects ${\cal I}^{+}$.
 One could regard
them as surfaces of constant time.

\subsection{Black Hole on a Surface}
A {\it black hole on the surface } $\Sigma (\tau)$ is
 a connected component of the set
$$
B(\tau)=\Sigma (\tau)- J^{-}({\cal I}^{+},\bar{M}).$$

One is interested primarily in
black holes which form from an initially
non-singular state. Such a state
can be described by using the partial
Cauchy surface $\Sigma$ which has an {\it asymptotically simple past},
i.e.
the causal past $J^{-}({\Sigma})$ is isometric to the region
 $J^{-}(\cal I)$ of some asymptotically
 simple and empty space-time with
 a Cauchy surface ${\cal I}$. Then $\Sigma$ has the topology $R^{3}$.

\subsection{ Boundary conditions in the path integral representation }
One is interested primarily in
black holes which form from an initially
non-singular state. Such a state
can be described by using the partial
Cauchy surface $\Sigma$ which has an {\it asymptotically simple past},
i.e.
the causal past $J^{-}({\Sigma})$ is isometric to the region
 $J^{-}(\cal I)$ of some asymptotically
 simple and empty space-time with
 a Cauchy surface ${\cal I}$. Then $\Sigma$ has the topology $R^{3}$.

In the path integral representation considered (\ref{FI}) we suppose that we deal with
 a set of space-times
$\{(M,g_{\mu\nu})\}$ which are weakly asymptotically simple and empty
and  strongly asymptotically
predictable. We take into account only such $(M,g_{\mu\nu})$ that contains
$\Sigma ' $  and $\Sigma ''$ so that
\begin{itemize}
\item
$\Sigma ' $ is a partial Cauchy surface with asymptotically simple past,
$\Sigma ' \sim R^{3}$.
\item
$\Sigma ''=\Sigma (\tau '') $ contains a black hole, i.e. $\Sigma '' -$
$J^{-}({\cal I} ^{+}, \bar{M})$ is nonempty.
\end{itemize}

In particular, if one has the
condition $\Sigma ' \cap \bar {J}^{-}({\cal I})$
is homeomorphic to $R^{3}$ (an open set with compact closure)
then $\Sigma ''$ also satisfies this condition.

\end{document}